\documentclass[12pt, colors]{iopart}

\usepackage{iopams}
\usepackage{bbold}
\usepackage{color}
\usepackage{xcolor}
\usepackage{graphicx}
\usepackage{hyperref}

\begin{document}

\title[Author guidelines for IOP Publishing journals in  \LaTeXe] 
{Fluctuation-Dissipation Relations in the absence of Detailed Balance:  formalism and applications to Active Matter}

\author{Sara Dal Cengio$^{1}$, Demian Levis$^{1,2}$, Ignacio Pagonabarraga$^{1,2,3}$}

\address{$^{1}$Departament de F\'isica de la Mat\`eria Condensada, Universitat de Barcelona, Mart\'i i Franqu\`es 1, E08028 Barcelona, Spain \\$^{2}$UBICS University of Barcelona Institute of Complex Systems, Mart\'i i Franqu\`es 1, E08028 Barcelona, Spain \\$^{3}$
Centre Europ\'een de Calcul Atomique et Mol\'eculaire (CECAM) , \'Ecole Polytechnique
F\'ed\'erale de Lasuanne (EPFL), Batochime, Avenue Forel 2, Lausanne, Switzerland}
\ead{sdalcengio@ub.edu}
\vspace{10pt}
\begin{indented}
\item[]February 2020
\end{indented}

\begin{abstract}
We present a comprehensive study about the relationship between  the way Detailed Balance is broken in non-equilibrium systems and the resulting violations of the Fluctuation-Dissipation Theorem. Starting from stochastic dynamics with both odd and even variables under Time-Reversal, 
 we exploit the relation between entropy production and the breakdown of Detailed Balance to establish general constraints on the non-equilibrium steady-states (NESS), which relate the non-equilibrium character of the dynamics with symmetry properties of the NESS distribution. This provides a direct route to derive extended Fluctuation-Dissipation Relations,  expressing the linear response function in terms of NESS correlations. 
  Such framework provides a unified way to understand the departure from equilibrium of active systems  and its linear response.  We then consider two paradigmatic models of interacting self-propelled particles, namely Active Brownian Particles (ABP) and Active Ornstein-Uhlenbeck Particles (AOUP). 
 We analyze the non-equilibrium character of these systems (also within a Markov  and a Chapman-Enskog approximation) and derive extended Fluctuation-Dissipation Relations for them, clarifying which features of these active model systems are genuinely non-equilibrium.
 \end{abstract}
\date{\today}

%
%
\submitto{\JSTAT}
%
%
%

\newpage

\tableofcontents

\newpage

\section{Introduction}\label{sec:1}

The Fluctuation-Dissipation Theorem (FDT) relates the correlations of spontaneous fluctuations, to the fluctuations induced by external stimuli \cite{KuboBook}. In practice, it allows to probe the response to external fields by analyzing the corresponding time-dependent equilibrium fluctuations, either in experiments or in simulations. For instance, it allows to  infer transport or mechanical properties of soft materials from light scattering without ever perturbing them \cite{brown1993dynamic, CipellettiRev}.
The FDT plays a particular important role in statistical mechanics as it is among the very rare general results in non-equilibrium,  although near to, conditions. 
It is valid for any equilibrium system (both in the classical and quantum realm) gently driven out-of-equilibrium by a small perturbation. 
Accordingly to the FDT, the response of an observable $A$ at time $t$ to a perturbation  $h$ , applied at time $s$, and causing the change in the energy of the system $E \to E - h(s)B$, is determined by an equilibrium correlation function as
\begin{equation}\label{eq:Kubo}
\left.\frac{ \delta\langle A(t)\rangle}{\delta h(s)}\right|_{h\to 0}=R_{A}(t, s) = \beta \frac{\partial}{\partial s} \langle A(t) B(s) \rangle_{\rm eq}, \quad t > s
\end{equation}
where $R_{A}$ is the response function of the observable $A$ reacting to a perturbation conjugated to $B$ and $\langle A(t) B(s) \rangle_{\rm eq}$ is the equilibrium correlation function between these two latter observables at temperature $\beta^{-1} = k_B T$ \footnote{We consider, without loss of generality, observables with zero mean.}.



 In its general formulation above, the FDT was first derived in the context of Hamiltonian mechanics \cite{doi:10.1143/JPSJ.12.570,kuboFDT}, where the dynamics is specified via a Liouville operator and the equations of motion are invariant under Time-Reversal. 
It has later been extended to stochastic descriptions \cite{HANGGI1982207} which rely on the hypothesis of scale separation between the system of interest and the bath, the latter being a collection of  (many) degrees of freedom with fast relaxation to equilibrium. Once such distinction is settled, dissipative and noisy terms enter into the equations of motion. As a result, the latter are no longer invariant under Time-Reversal. Nevertheless a footprint of reversibility holds at the stochastic description level under the name of Detailed Balance (DB)\cite{Risken, Graham}. As long as DB is guaranteed, the FDT holds, both in thermal and athermal states \cite{Agarwal, Haken1969}.

For systems breaking DB,  relentlessly evolving far-from-equilibrium, the FDT is no longer justified. The question of whether a similar relation as eq.~(\ref{eq:Kubo})  can be derived in this case, has been the focus of a great deal of research efforts over the last decades. In particular, several extended Fluctuation-Dissipation Relations (FDR) have been derived, using different approaches, for systems in non-equilibrium steady-states (NESS). 
However, contrary to equilibrium states, no universal relation such as eq. (\ref{eq:Kubo}) exists for NESS. The establishment of a general  extended FDR with the features of the equilibrium FDT, remains a central challenge towards the construction of a general framework to deal with non-equilibrium systems. 
In the context of stochastic dynamics,
 extended FDR for NESS have mostly focused on overdamped descriptions \cite{PhysRevLett.95.130602, SpeckSeifert, ProstJoanny, BaiesiMaesPRL, Chetrite2009,  SeifertSpeck, Kurchan_1998}. We refer to \cite{MarconiReview, BaiesiMaesNJP, SarracinoRev}  for recent reviews on the topic. 

Among the variety of non-equilibrium systems, living matter constitutes a particularly interesting  class.  From a physics viewpoint, it can be considered as active matter: systems composed of interacting units  - be it a cell, a molecular motor, a auto-catalytic colloid -  capable of extracting energy from their environment to perform some task and, typically (as in the cases we consider here), self-propel. In contrast with passive systems relaxing towards NESS, which are driven out-of-equilibrium by external global means (usually through their boundaries),  active systems break DB at the level of each of its constituents, defining a fundamentally different class of non-equilibrium systems \cite{CatesRev, RevModPhys.88.045006}.

 A renewed interest in the characterization of NESS comes indeed from active matter physics.  
 The possibility to extend equilibrium-like concepts to characterize active matter, in particular their NESS, has been the focus of intense efforts over the past decade. Most of our general understanding of  such  fundamental aspects of active matter has been gained through the detailed investigation of  simple models of self-propelled particles, such as the  Active Brownian Particles (ABP) \cite{LutzABP, FilyMarchetti, PhysRevLett.110.055701, PhysRevLett.121.098003} and Active Ornstein-Uhlenbeck Particles (AOUP)  \cite{szamel2014self, MatteoMaggi2014, MarconiMaggi, FodorPRL, StefanoPRX, Bonilla} models that we consider here. 
Quantities such as the pressure or chemical potential have been defined for model active systems and exploited to characterize their phase behavior and the nature of the (non-equilibrium) phase transitions they exhibit  \cite{Brady2014, mallory2014anomalous, Ginot, SolonNature, winkler2015virial,wittkowski2014scalar,  C3SM52813H, C7SM01504F, paliwalChemical,  steffenoni2017microscopic, solon2018generalized, PhysRevLett.121.098003, PhysRevLett.123.268002}.  

Attempts to extend the FDT to characterize the linear response of active systems has been limited to specific cases or regimes, mostly considering activity as a small parameter.
In \cite{SharmaBrader2016, ThomasABPMCT}, activity is treated as the perturbation on an otherwise equilibrium state, while in \cite{FodorPRL} a FDR is obtained in a small activity regime for which the dynamics of the system fulfills DB. In both cases, the reference state that is perturbed is not a genuine NESS: in the first case, it is an equilibrium state with Boltzmann statistics, while in the latter an effective equilibrium state with a generalized potential.  The fundamental difficulties arising from the violation of DB are therefore bypassed.  The linear response beyond such small activity limit has been analyzed  for a single active particle in \cite{Caprini2018}. In \cite{SolonResponse}, response functions were obtained beyond such limit regimes, although they are not  written in terms of NESS time-correlation functions, as one wills for establishing FDR, but as weighted averages (in the spirit of Malliavin weight sampling \cite{WarrenAllenPRL}). 
Another strategy consists in systematically quantifying the violations of the FDT through an effective temperature \cite{cugliandolo2011effective, loi2008effective, szamel2014self, LevisBerthier, Yair2011effective, LeticiaIsabella, Wylie, BennoTeff, SzamelNew, Isabella}. While this approach provides useful insights into the dynamics of NESS, it does not carry the same piece of information as a FDR, i.e. a generic way to asses the  response function of an active system in terms of the steady-state fluctuations of measurable observables. 
Activity results on transport phenomena which are impossible in equilibrium passive systems, as recently observed in experiments involving biological microorganisms \cite{diLeonardoWheel, Sokolov969, BacteriaClement, RafaiPRL2010} as well as artificial phoretic motors \cite{doi:10.1002/anie.200600060, PhysRevLett.105.088304, Palacci936, PhysRevLett.108.268303}. A key step towards the fundamental understanding of active materials  is to characterize transport coefficients and establish extended Green-Kubo expressions resulting from the FDR. 

Here we address the question of how systems interacting active particles respond to an external small perturbation. 
Although the non-equilibrium nature of active systems is intrinsically different from the one of passive driven systems, as for the construction of a   linear response theory, the fundamental difficulty to be tackled in both cases is the breakdown of DB.
We first establish a general constraint on the NESS to be fulfilled by \emph{any} Markovian dynamics, fulfilling or not DB. The framework and results obtained apply to  both systems with only even variables and even and odd variables under Time-Reversal, such as overdamped and underdamped Langevin processes.  Such constraint on the NESS  stands for a relation between the nature of the non-equilibrium fluxes and the symmetry (under Time-Reversal) of the NESS distribution. We then derive an extended FDR for stochastic dynamics breaking DB.  
We finally turn to the application of these general results to archetypical  models of active particles: Active Brownian Particles (ABP)  and Active Ornstein-Uhlenbeck particles (AOUP). For ABP we consider the non-interacting limit and an effective equilibrium regime resulting from a Markovian approximation as discussed in \cite{SaraPRL}.  For AOUP we derive a genuine, although approximated, non-equilibrium FDR  unveiling the interplay between activity and interactions. We discuss in detail the specificities of AOUP as compared with ABP as well as the   different approximation schemes used in the literature to deal with many-body effects.


The paper is organized as follows. In section \ref{sec:2} we establish the general framework and notation used throughout the paper. Section \ref{sec:3}  recalls some general aspects of equilibrium dynamics that are important to clarify before moving to  non-equilibrium dynamics. A reader familiar with the formalism of stochastic processes may directly move to section \ref{sec:4}, where general aspects of non-equilibrium dynamics are discussed: We derive a general expression for the generator of the time-reversed dynamics and connect it to the concept of entropy production,  allowing the derivation of general constraints  a non-equilibrium stationary measure must fulfill. An extended FDR valid for systems breaking DB is then derived and discussed.  Section \ref{sec:5} is dedicated to the application of these results to simple models of self-propelled particles (ABP and AOUP). Section \ref{sec:6} contains our conclusions and final remarks.

\section{Stochastic dynamics: general aspects and definitions}\label{sec:2}
\subsection{Fokker-Planck equation}
Our starting point is a generic system with $N$ dynamic variables $\boldsymbol{\Gamma} \equiv\{ \Gamma_i \}_{i=1}^{N}$ defined on a manifold $\mathcal{M} \subset \mathbb{R}^N$. We introduce a probability distribution $\Psi $ which assigns  $\Psi(\boldsymbol{\Gamma}, t)$ to any point $\boldsymbol{\Gamma}\in \mathcal{M} $ at a time $t$. The implicit assumption is the requirement for $\Psi$ to be smooth enough for its partial derivatives to exist. 

Generically the time evolution of $\Psi(\boldsymbol{\Gamma}, t)$ is described by a generator $\Omega_0 (\boldsymbol{\Gamma})$:
\begin{equation}\label{eq:generic}
\partial_t \Psi(\boldsymbol{\Gamma}, t ) = \Omega_0 (\boldsymbol{\Gamma})\Psi(\boldsymbol{\Gamma}, t )
\end{equation}
together with an appropriate initial condition $\Psi(\boldsymbol{\Gamma}, 0)$. Formal integration of eq.~(\ref{eq:generic}) leads  to $\Psi(\boldsymbol{\Gamma}, t ) = e^{\Omega_0 t} \Psi(\boldsymbol{\Gamma}, 0) $. We denote $\Psi_0$ the steady-state solution of the dynamics above, meaning 
\begin{equation}\label{eq:stationarity}
 \Omega_0 \Psi_0=0 \, .
\end{equation}
Up to here, we did not need to specify  the nature of the dynamics. We shall now focus on stochastic  dynamics (although the following formalism  could be extended to, say, Hamiltonian dynamics). In that case, an extra assumption is needed to ensure that eq.~(\ref{eq:generic}) is fully determined by the initial condition $\Psi(\boldsymbol{\Gamma},0)$ \textit{i.e.} the requirement of markovianity for  $\{ \Gamma_i \}$ \cite{vanKampen1989}.
Whenever this assumption is met, the generator in eq.~(\ref{eq:generic}) has the so-called Fokker-Planck form:
\begin{equation}\label{eq:evolutionoperator}
\Omega_0(\boldsymbol{\Gamma}) = \sum_i \left( -\partial_i \mathcal{A}_i(\boldsymbol{\Gamma}) + \sum_j \partial_i \partial_j \mathcal{B}_{ij}(\boldsymbol{\Gamma}) \right)
\end{equation}
where $\partial_i \equiv \partial/\partial \Gamma_i$, $\mathcal{A}\equiv \{\mathcal{A}_i \}_{i=1}^{N}$ is the drift vector and $\mathcal{B} \equiv \{ \mathcal{B}_{ij} \}_{i,j = 1}^{N}$ is the $N\times N$ diffusion matrix. 
 In the following, unless explicitly stated otherwise, we will take $\mathcal{B}$ to be invertible and diagonal with constant entries, such that $\mathcal{B}_{ij} \equiv D_i  \delta_{ij}$.


The dynamics is fully specified by the knowledge of $\Psi(\boldsymbol{\Gamma}, t) $ or, equivalently, by the knowledge of the conditional probability density $P(\boldsymbol{\Gamma}, t |\boldsymbol{\Gamma}_0, t_0)$ defined as  the probability to be in $\boldsymbol{\Gamma}$ at time  $t$ given the configuration $\boldsymbol{\Gamma}_0$ at time $t_0$. Eq.~(\ref{eq:generic}) can be recast in terms of $P(\boldsymbol{\Gamma}, t| \boldsymbol{\Gamma}_0, t_0)$ as
\begin{equation}\label{eq:forward}
\partial_t P(\boldsymbol{\Gamma}, t | \boldsymbol{\Gamma}_0, t_0) = \Omega_0(\boldsymbol{\Gamma}) P(\boldsymbol{\Gamma}, t | \boldsymbol{\Gamma}_0, t_0)
\end{equation}
which is often called \textit{forward} equation to distinguish it from the \textit{backward} equation:
\begin{equation}\label{eq:backward}
\partial_{t_0} P(\boldsymbol{\Gamma}, t | \boldsymbol{\Gamma}_0, t_0) = -\Omega_0^{\dagger}(\boldsymbol{\Gamma}_0) P(\boldsymbol{\Gamma}, t | \boldsymbol{\Gamma}_0, t_0)
\end{equation}
where $
\Omega^{\dagger}_0(\boldsymbol{\Gamma}) = \sum_i  \mathcal{A}_i(\boldsymbol{\Gamma})\partial_i + D_i\partial_i^2 
$ is the adjoint operator of $\Omega_0$.
The main difference between the two equations is the set of variables that we hold fix. In eq.~(\ref{eq:generic}) we fix  the \textit{initial} condition at time $t_0$ and we look at the evolution for $t > t_0$. In eq.~(\ref{eq:backward}), instead, we fix the \textit{final} condition at time $t$ and we look at the evolution for $t_0 < t$. This remark will show its relevance when characterizing the departure from equilibrium in systems breaking DB. 

\subsection{Symmetry aspects under Time-Reversal}

We shall distinguish the dynamic variables $\Gamma_i$ according to their parity under  Time-Reversal  
\begin{equation}
 \mathcal{T}: \boldsymbol{\Gamma} \in \mathcal{M} \mapsto \boldsymbol{\varepsilon}\boldsymbol{\Gamma} \equiv \{ \varepsilon_i \Gamma_i \} \in \mathcal{M},\\ \varepsilon_i = \pm 1 .
 \end{equation}
 Variables $\Gamma_i$ for which $\varepsilon_i = 1$ are said even under Time-Reversal and variables for which $\varepsilon_i = -1$ are said to be odd. For instance, if one has in mind the dynamics of a particle in phase space, $\boldsymbol{\Gamma}=(x,p) $ and
 \begin{equation}
 \mathcal{T}: (r,\,p)  \mapsto  (r,\,-p) \, .
 \end{equation}
The position variable $r$ is even while momentum $p$ is odd. 

The Fokker-Planck equation stands for the conservation of the probability density and can thus be written in terms of a probability flux $\boldsymbol{J}\equiv \{J_i \}_{i=1}^{N}$ as
\begin{eqnarray}\label{eq:genericFP}
\Omega_0 \Psi (\boldsymbol{\Gamma}, t) &=&-  \boldsymbol{\nabla} \cdot \boldsymbol{J}(\boldsymbol{\Gamma}, t)=\partial_t \Psi (\boldsymbol{\Gamma}, t)  \\
J_i (\boldsymbol{\Gamma}, t) &= & \mathcal{A}_i(\boldsymbol{\Gamma}) \Psi(\boldsymbol{\Gamma}, t) - D_i\partial_i \Psi(\boldsymbol{\Gamma}, t) \label{eq:flux}
\end{eqnarray}
where $\boldsymbol{\nabla} \equiv \{ \partial_i  \}$.
Since we allow $\{\Gamma_i \}$ to be either even or odd under Time-Reversal, we  can decompose the drift vector in  a \emph{reversible}  and an \emph{irreversible} part, $\mathcal{A}=\mathcal{A}^{\rm rev}+\mathcal{A}^{\rm irr}$,  defined as
\begin{eqnarray}\label{eq:Arev}
\mathcal{A}^{\rm rev}_i (\boldsymbol{\Gamma}) &\equiv& \frac{1}{2} \left[ \mathcal{A}_i (\boldsymbol{\Gamma}) - \varepsilon_i \mathcal{A}_i  (\boldsymbol{\varepsilon}\boldsymbol{\Gamma}) \right]  \\ 
\mathcal{A}^{\rm irr}_i(\boldsymbol{\Gamma})&\equiv &\label{eq:Airr} \frac{1}{2} \left[ \mathcal{A}_i  (\boldsymbol{\Gamma})+ \varepsilon_i \mathcal{A}_i (\boldsymbol{\varepsilon}\boldsymbol{\Gamma}) \right]
\end{eqnarray}
which, under Time-Reversal transform as
\begin{equation}\label{eq:TR_Airr}
\mathcal{A}^{\rm rev}_i (\boldsymbol{\varepsilon \boldsymbol{\Gamma}}) =  - \varepsilon _i  \mathcal{A}^{\rm rev}_i (\boldsymbol{ \boldsymbol{\Gamma}}),\\ \mathcal{A}^{\rm irr}_i(\boldsymbol{\varepsilon}\boldsymbol{\Gamma}) =  \varepsilon_i  \mathcal{A}^{\rm irr}_i(\boldsymbol{\Gamma}) \, .
\end{equation}
We thus  identify  two  distinct contributions to the total probability flux $J_i(\boldsymbol{\Gamma}, t) = J_i^{\rm rev} (\boldsymbol{\Gamma}, t)+ J_i^{\rm irr}(\boldsymbol{\Gamma}, t) $, where
\begin{eqnarray}
J_i^{\rm rev} (\boldsymbol{\Gamma}, t) &=&  \mathcal{A}^{\rm rev}_i(\boldsymbol{\Gamma}) \Psi(\boldsymbol{\Gamma}, t)  \label{eq:fluxrev}
\\
J_i^{\rm irr} (\boldsymbol{\Gamma}, t) &=& \mathcal{A}^{\rm irr}_i(\boldsymbol{\Gamma})
 \Psi(\boldsymbol{\Gamma}, t)  -  D_i\partial_i \Psi(\boldsymbol{\Gamma}, t) \, .\label{eq:fluxirrev}
\end{eqnarray}
We denote the steady-state flux $\boldsymbol{J}_0$ and  define  the   \emph{phase-space  velocity} as
\begin{equation}
 \boldsymbol{\mathcal{V}} (\boldsymbol{\Gamma}) \equiv\boldsymbol{J}_0(\boldsymbol{\Gamma})/\Psi_0 (\boldsymbol{\Gamma}) =\{ \mathcal{V}^{\rm rev}_i+\mathcal{V}^{\rm irr}_i\}_{i=1}^{N}  .
\end{equation}
[In the following,  we also report  its time-dependence $\boldsymbol{\mathcal{V}}(\boldsymbol{\Gamma}, t) \equiv \boldsymbol{J}(\boldsymbol{\Gamma}, t) /\Psi(\boldsymbol{\Gamma}, t)$].
The  decomposition of the probability flux into two contributions with different symmetry under Time-Reversal  will play a central role in the following treatment  \cite{PhysRevLett.81.3063, Qian2002, Wang12271}.  All the dissipative terms are embedded in eq.~(\ref{eq:fluxirrev}). 

To illustrate the definitions above, let us consider a simple example: a  Brownian particle, moving in one dimension, at position ${x}(t)$ and momentum ${p}(t)$ at time $t$, i.e. $\boldsymbol{\Gamma}=({x}, {p})$, described by the following Langevin equation:
\begin{equation}\label{eq:underdamped}
 \dot{{x}}(t)={p}(t)\\
\dot{{p}}(t)=-U'(x)-\gamma{p}(t)+\sqrt{2\gamma  k_BT}{\xi}(t)
\end{equation}
where $-U'(x)$ is the total force exerted on the particle, which can either come from inter-particle interactions or an external field, $\gamma$ is the drag coefficient and ${\xi}(t)$ is a Gaussian white noise of zero mean and unit variance. Here and in the rest of the paper we set the mass $m\equiv 1$. We  consider here the one dimensional case for simplicity, though the extension to higher dimension is straightforward. 
The last two terms of the right hand side account for the coupling of the particle with a thermal bath at temperature $T$, source of noise and dissipation. In equilibrium, the FDT constraints the amplitude of the noise and dissipation to be related, as made apparent in the equation above. 
The latter Langevin equation, can be equivalently written as the Fokker-Planck equation, with  drift  vector
\begin{eqnarray}\label{eq:Aunderdamped}
\mathcal{A}=\left[\begin{array}{c}\label{eq:Aunderdamped}
\mathcal{A}_{{x}}={p}(t) \\
\mathcal{A}_{{p}}=-U'(x)-\gamma {p}(t)
\end{array}\right]
\end{eqnarray}
and diffusion matrix
\begin{eqnarray}\label{eq:coefficientsunderdamped}
\mathcal{B}=\left[\begin{array}{cc}0 & 0 \\
0 & \gamma k_BT\end{array}\right] \, .\label{eq:Bsingular}
\end{eqnarray}
Under Time-Reversal, the dynamic variables transform as
\begin{eqnarray}
\boldsymbol{\varepsilon}\boldsymbol{\Gamma}=\boldsymbol{\varepsilon}\left[\begin{array}{c}x \\p\end{array}\right]=\left[\begin{array}{c}x \\-p\end{array}\right] \, .
\end{eqnarray}
We now apply the definition of the reversible and irreversible drift vectors Eqs.~(\ref{eq:Arev}-\ref{eq:Airr}) and find
\begin{eqnarray}
\mathcal{A}^{\rm rev} =\left[\begin{array}{c}
{p}(t) \\
-U'(x)
\end{array}\right]\, \, , \, \,\mathcal{A}^{\rm irr} =\left[\begin{array}{c}0\\
-\gamma{p}(t) 
\end{array}\right] \, . \\
\end{eqnarray}
In the overdamped limit, the Brownian particle can be described by 
\begin{equation}\label{eq:overdamped}
\dot{{x}}(t)=-\mu U'(x)+\sqrt{2\mu k_BT}{\xi}(t)
\end{equation}
where $\mu=1/\gamma$. 
The drift and diffusion vector thus read
 \begin{equation}
\mathcal{A}=-\mu U'(x)=\mathcal{A}^{\rm irr} \,, \\
\mathcal{B}=\mu k_B T=D_x\,  .
\end{equation}
In this case, the only dynamic variable is ${x}$, which is even.  The absence of odd variables implies the absence of reversible fluxes, and thus $\mathcal{A}^{\rm rev}$ is identically zero.

It is worth at this stage to make  a few remarks. For the underdamped system, $\mathcal{B}$ is non-invertible. This will have a consequence on the determination of the steady-state distribution fulfilling DB as shown in the next sub-section; see therein for more details. 
Finally, in the absence of dissipation (and diffusion) the generator $\Omega_0$ would be identified with the Liouville operator. In that case, the irreversible part of the flux would vanish and the motion would be purely reversible, as expected from Time-Reversal symmetry of the `microscopic' Hamilton equations of motion. 

\subsection{Observables}
In this section we fix some notations and definitions that will be used in the following.  

Physical observables are represented by real functions acting on $\mathcal{M}$,  such that $A: \boldsymbol{\Gamma} \in \mathcal{M}  \mapsto A(\boldsymbol{\Gamma}) \in \mathbb{R}$.  Their  steady-state average  is defined as
\begin{equation}\label{eq:ensembleschod_ss}
\langle A \rangle_0= \int_{\mathcal{M}} d\boldsymbol{\Gamma} A(\boldsymbol{\Gamma}) \Psi_0(\boldsymbol{\Gamma}) 
\end{equation}
and the ensemble average at time $t$ is defined as
\begin{equation}\label{eq:ensembleschod_t}
\langle A \rangle_t = \int_{\mathcal{M}} d\boldsymbol{\Gamma} A(\boldsymbol{\Gamma}) \Psi(\boldsymbol{\Gamma}, t) = \int_{\mathcal{M}} d\boldsymbol{\Gamma} A(\boldsymbol{\Gamma}) e^{\Omega_0 t} \Psi(\boldsymbol{\Gamma}, 0) \, .
\end{equation}
The time evolution may be given  to the observables (instead of the probability distribution) using the adjoint of the Fokker-Planck generator
\begin{equation}\label{eq:ensembleheis}
\langle A\rangle_t = \int_{\mathcal{M}} d\boldsymbol{\Gamma} e^{\Omega_0^{\dagger}t}A(\boldsymbol{\Gamma})  \Psi(\boldsymbol{\Gamma}, 0) \equiv\int_{\mathcal{M}} d\boldsymbol{\Gamma}A(t) \Psi(\boldsymbol{\Gamma}, 0) .
\end{equation}
Both expressions of  $\langle A \rangle_t $ are fully equivalent and  are respectively referred to as the Schr\"{o}dinger and Heisenberg representation, by analogy with quantum mechanics \cite{10.2307/j.ctt24h99q}. In the Schr\"{o}dinger the time dependence is encoded in the probability density (analogous to the wave-function), while in the Heisenberg representation, the time dependence is encoded in the observables, which are now explicitly time-dependent and are evolved by the adjoint of the evolution operator (analogous to Hermitian operators acting on a Hilbert space).

The  function $R_{A}(t,s)$ encodes the linear response of an observable $A$, due to a  perturbation $h$, applied at $t=0$, conjugated to an observable $B$ which results in a change of generator $\Omega_0\to\Omega=\Omega_0+\Omega_{{ext}}$. It is defined as 
\begin{equation}
\langle A \rangle_t- \langle A\rangle_0=\int_0^t ds R_{A}(t, s) h(s) +O(h^2)\,, \quad t > s.
\end{equation}
Then, writing the response function  $R_{A}$ as given by the equilibrium FDT eq.  \ref{eq:Kubo},  considering a constant perturbation $h$ and integrating by parts,  we obtain
\begin{equation}
\langle A \rangle_t- \langle A\rangle_0=h\beta[\langle A(t)B(t)\rangle-\langle A(t)B(0)\rangle]
\end{equation}
In its integrated version, the equilibrium FDT reduces to a simple linear relation  between the integrated response and its conjugated correlation function.



\section{A preamble: Equilibrium dynamics}\label{sec:3}
\subsection{Detailed Balance}
\begin{figure}
\includegraphics[width=\textwidth]{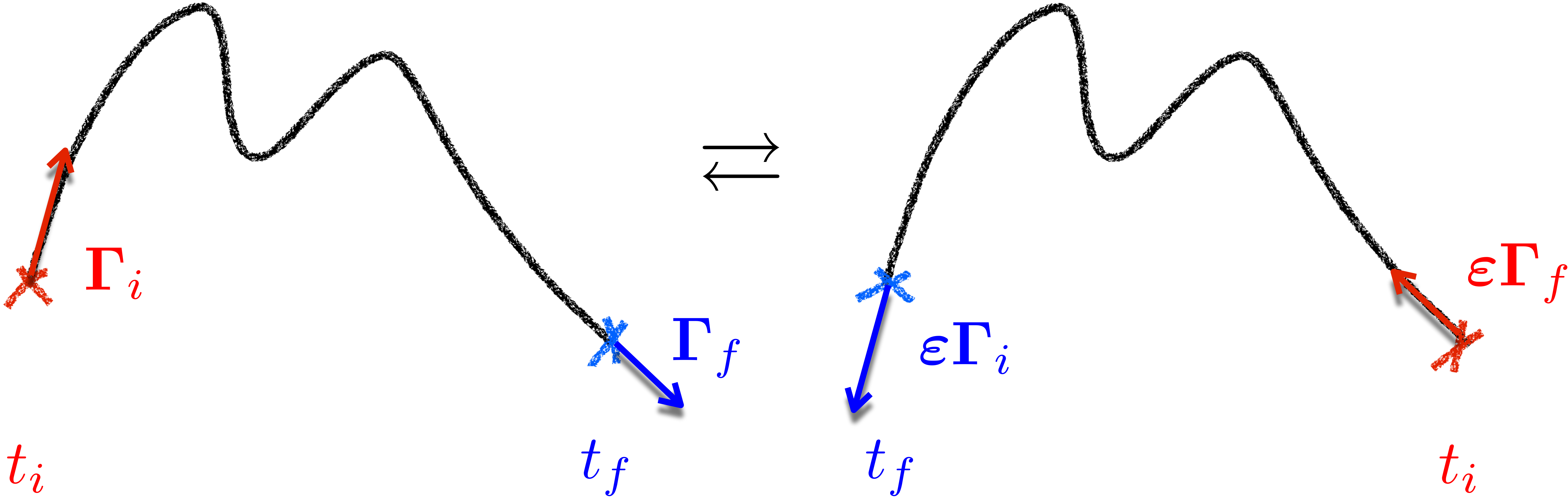}
\caption{Illustration of a dynamics verifying Detailed Balance: the probability to transit  from an initial steady-state $\boldsymbol{\Gamma}_i$, to a final steady-state  $\boldsymbol{\Gamma}_f$, must be equal to the probability of the reverse transition, meaning, the probability of the transition from an initial steady-state $\boldsymbol{\epsilon}\boldsymbol{\Gamma}_f$ (the time-reversed version of $\boldsymbol{\Gamma}_f$), to  $\boldsymbol{\epsilon}\boldsymbol{\Gamma}_i$. }
\label{fig:DB}
\end{figure}

Whether we interpret a stochastic dynamics as deriving from an underlying microscopic description fulfilling the laws of classical or quantum mechanics or not, DB constitutes the key symmetry of equilibrium. A   system  
 is said to satisfy DB if, at stationarity,  any microscopic process is balanced by the reversed one. It can thus be formally written as  \
 \begin{equation}\label{eq:DBconditional}
P(\boldsymbol{\Gamma}_f, t_f |\boldsymbol{\Gamma}_i, t_i) \Psi_0(\boldsymbol{\Gamma}_i) = P( \boldsymbol{\varepsilon} \, \boldsymbol{\Gamma}_i, t_f |\boldsymbol{\varepsilon} \boldsymbol{\Gamma}_f, t_i) \Psi_0( \boldsymbol{\varepsilon} \boldsymbol{\Gamma}_f)\,
\end{equation}
for any pair of states $(\boldsymbol{\Gamma}_i, \boldsymbol{\Gamma}_f)$ and at any times $(t_i, t_f)$ (see cartoon Fig. \ref{fig:DB}).
%
%
By setting $t_i = t_f$ in eq.~(\ref{eq:DBconditional}) we get:
\begin{equation}\label{eq:evenPsi}
\Psi_0(\boldsymbol{\Gamma}) = \Psi_0(\boldsymbol{\varepsilon} \boldsymbol{\Gamma}),\, \quad \forall\, \boldsymbol{\Gamma}\in\mathcal{M} \, .
\end{equation}
It follows that the mean value of any  \textit{current-like} observable,  \textit{i.e.} $A(\boldsymbol{\varepsilon} \boldsymbol{\Gamma}) = -A(\boldsymbol{\Gamma})$ must be zero if DB is fulfilled \footnote{Note that here we are referring to physical currents and not to the probability current $J_i$ in eq.~(\ref{eq:genericFP}). Indeed, eq.~(\ref{eq:evenPsi}) implies $\langle A \rangle_0=\int d\boldsymbol{\Gamma} A(\boldsymbol{\boldsymbol{\varepsilon}\Gamma}) \Psi_0(\boldsymbol{\varepsilon}\boldsymbol{\Gamma})$=
$-\int d\boldsymbol{\Gamma} A(\boldsymbol{\Gamma}) \Psi_0(\boldsymbol{\Gamma})=-\langle A \rangle_0$ if $A(\boldsymbol{\varepsilon} \boldsymbol{\Gamma}) = -A(\boldsymbol{\Gamma})$.}. 
Rather than being a symmetry at the level of single trajectories (as for the microscopic description), DB is formulated in eq.~(\ref{eq:evenPsi}) as a symmetry property of the steady-state distribution $\Psi_0(\boldsymbol{\Gamma}$).
Actually, a necessary and sufficient condition for DB to hold is the absence of irreversible fluxes in steady conditions \cite{GardinerBook}:
\begin{eqnarray}\label{eq:gardinercondition}
{\rm Detailed\, Balance \,} \Leftrightarrow \boldsymbol{J}_0^{\rm irr}=0
\end{eqnarray}
where $\boldsymbol{J}_0^{\rm irr}=\{\mathcal{A}_i^{\rm irr}(\boldsymbol{\Gamma})\Psi_0(\boldsymbol{\Gamma})-D_i\partial_i\Psi_0(\boldsymbol{\Gamma})\}_{i=1}^{N}$. 
This means that, reversible steady-state fluxes are not constrained by DB. To illustrate this aspect let us go back to the example of the Brownian particle. In the absence of even  variables, i.e. in the overdamped regime, DB corresponds precisely to the absence of steady-state fluxes (since $J^{\rm  rev} = 0$). However, in the presence of odd (momentum-like) variables as in the underdamped description, reversible steady-state fluxes can be present in a system fulfilling DB   (although  physical currents must have all zero ensemble averages, see eq.~(\ref{eq:evenPsi})) \cite{TailleurKurchan2006, JorgeLectures}. 

%
The absence of irreversible fluxes can be rewritten as
\begin{equation}\label{eq:almostDB}
\partial_i \log \Psi_0(\boldsymbol{\Gamma}) =   D_i^{-1} \mathcal{A}_i^{\rm irr} (\boldsymbol{\Gamma}), 
\end{equation}
which in turn imposes the so-called 'thermodynamic  curvature' (the curl in $N$ dimensions)
 \cite{polettini2013dice} of the irreversible drift to vanish: 
\begin{equation}\label{eq:thermocurv}
D_i^{-1} \partial_j\mathcal{A}_i^{\rm irr}(\boldsymbol{\Gamma})-  D_j^{-1} \partial_i\mathcal{A}_j^{\rm irr}(\boldsymbol{\Gamma})= 0 \, .
\end{equation}
The latter expression provides an alternative definition of DB in terms of geometrical properties of the drift and diffusion terms. The advantage of eq.~(\ref{eq:thermocurv}) over the definition in Eqs.(\ref{eq:gardinercondition}-\ref{eq:almostDB}) is that it allows to verify if a dynamics fulfills or not DB without the need of finding $\Psi_0(\boldsymbol{\Gamma})$. The constraint imposed by DB on the steady-state currents provides a natural route to explicitly derive a steady solution. Whenever eq.~(\ref{eq:thermocurv}) is satisfied, one can derive a steady solution by direct integration. No such a procedure exists if DB is broken, and no prescribed functional form of the multivariate Fokker-Planck equation can be derived in general. This is precisely the great advantage of equilibrium dynamics:  the steady-state can be solved just by quadrature, giving the  equilibrium distribution
\begin{equation}\label{eq:potentialcondition}
\Psi_0(\boldsymbol{\Gamma}) =\Psi_{\rm eq}(\boldsymbol{\Gamma}) = \mathcal{N} \exp \left[ \sum_i \int D_i^{-1}  \mathcal{A}_i^{\rm irr} d \Gamma_i \right]
\end{equation}
with $\mathcal{N}$ the normalization constant such that $\int d\boldsymbol{\Gamma} \Psi_0(\boldsymbol{\Gamma}) = 1$.

It is straightforward to apply eq.~(\ref{eq:potentialcondition}) to the case of an equilibrium Brownian particle in the overdamped limit, eq.~(\ref{eq:overdamped}). In this case the integral in eq.~(\ref{eq:potentialcondition}) gives the Boltzmann distribution
$\Psi_{\rm eq}(\boldsymbol{\Gamma}) = \mathcal{N} \exp \left[ -\beta U({x})\right]$. As we mentioned previously, a little care must be taken when applying eq.~(\ref{eq:potentialcondition}) to the underdamped particle of eq.~(\ref{eq:underdamped}) - since in this case, the diffusive matrix is non invertible, see eq. \ref{eq:coefficientsunderdamped}. The integral over the phase space in eq.~(\ref{eq:potentialcondition}) has to be carried on momenta only,  to find
\begin{equation}
 \Psi_{\rm eq} (\boldsymbol{\Gamma})\sim \exp\left[ -\beta {p}^2/2 + \Lambda({x})\right]
 \end{equation}
  with $\Lambda({x})$ a function of spatial coordinates to be determined by imposing stationarity (i.e. $\Omega_0\Psi_0=0$). In the presence of odd variables (underdamped case), the steady-state distribution is not fully determined by DB, which only constraints irreversible fluxes, but one has also to explicitly apply the evolution equation to specify the dependence on positions, which, as expected, results in $\Lambda=U$.



\subsection{The Fluctuation-Dissipation Theorem}


We focus now on the linear response of a system initially prepared in a steady-state.  At $t=0$ we apply an infinitesimal perturbation to the drift vector $ \mathcal{A} \to \mathcal{A} + \delta \mathcal{A}$. The evolution equation (\ref{eq:generic}) now reads:
 \begin{equation}\label{eq:perturbedFP}
 \partial_t \Psi(\boldsymbol{\Gamma}, t) =\Omega \Psi(\boldsymbol{\Gamma}, t) = \left[ \Omega_0(\boldsymbol{\Gamma}) + \Omega_{\rm ext}(\boldsymbol{\Gamma}) \right] \Psi(\boldsymbol{\Gamma}, t)
 \end{equation}
 where 
 \begin{equation}\label{eq:externaloperator}
 \Omega_{\rm ext}(\boldsymbol{\Gamma}) \Psi(\boldsymbol{\Gamma}, t) = - \boldsymbol{\nabla} \cdot \big[ \delta \mathcal{A}(\boldsymbol{\Gamma}) \Psi(\boldsymbol{\Gamma}, t) \big] \,  .
 \end{equation}
accounts for the perturbation.
 Since the generator $\Omega$ does not explicitly depend on time, we can write \cite{fuchs2005}:
 \begin{equation}
  \int_{0}^{t} \frac{d}{dt'} e^{\Omega t'} dt' = \int_{0}^{t} \Omega e^{\Omega t'} dt' = e^{\Omega t} -1\, .
 \end{equation}
We now use the latter expression into $\Psi(\boldsymbol{\Gamma}, t) =e^{\Omega t}\Psi(\boldsymbol{\Gamma}, 0)$ to obtain,  to first order in $\delta \mathcal{A}$, 
\begin{equation}
\Psi(\boldsymbol{\Gamma}, t) = \Psi_{0}(\boldsymbol{\Gamma}) + \int_{0}^{t} dt' e^{\Omega_0(\boldsymbol{\Gamma}) t'} \Omega_{\rm ext}(\boldsymbol{\Gamma}) \Psi_0(\boldsymbol{\Gamma}) \, .
\end{equation}
For an observable $A$, we then find the so-called Agarwal FDR  \cite{Agarwal}:
\begin{equation}\label{eq:agarwal}
 \langle A\rangle_t - \langle A \rangle_0 =\int_{0}^{t} ds \left\langle A(s) B (0) \right\rangle_0,\\ B(0) \equiv \frac{\Omega_{\rm ext} (\boldsymbol{\Gamma})\Psi_0(\boldsymbol{\Gamma})}{\Psi_0(\boldsymbol{\Gamma})}
\end{equation}
where $B$ is the observable conjugated to the perturbation $\delta \mathcal{A}$ and $\langle \cdot \rangle_0$ is the ensemble average weighted with the initial stationary distribution $\Psi_0$. Alternatively, using eq.~(\ref{eq:externaloperator}) into (\ref{eq:agarwal}) the integrated response can be expressed as:
\begin{equation}\label{eq:fdtvulpiani}
\langle A\rangle_t - \langle A \rangle_0 = -\int_0^{t} ds \langle A(s)\left[ \boldsymbol{\nabla} \cdot \delta \mathcal{A}+\delta \mathcal{A} \cdot \boldsymbol{\nabla} \log \Psi_0\right] (0)  \rangle_0
\end{equation}
which resembles the FDR as originally appeared in \cite{FALCIONI1990341} in the context of dynamical systems.
At this level, no equilibrium hypothesis has been made and, as such, Eqs. (\ref{eq:agarwal}-\ref{eq:fdtvulpiani}) remain valid also for NESS. 
For an equilibrium system fulfilling DB, i.e.  $\Psi_0(\boldsymbol{\Gamma}) = \Psi_{_{\rm eq}}(\boldsymbol{\Gamma})$, the conjugated  observable $B$ can be computed explicitly.

As an illustration, we consider again a one-dimensional overdamped  Brownian particle in contact with a thermal bath at temperature $T$ that we perturb by applying, at  $t=0$, a constant external force $h$. 
In this case  $\Omega_0 = \mu\partial_x [ k_BT \partial_x -  F ]$ with $F \equiv - U'(x)$ the conservative force acting on the particle and $\Omega_{_{\rm ext}} = -\mu h \partial_x $. 
From eq.~(\ref{eq:potentialcondition}) we immediately obtain the Boltzmann  distribution, and from eq.~(\ref{eq:agarwal})\begin{equation}
\langle  A \rangle_{t}  - \langle  A \rangle_{0} =- \mu\beta h \int_0^{t} ds \langle  A(s) F(0) \rangle_{0} \, .
\end{equation}
If we now replace $\mu F=\dot{x}-\sqrt{2\mu k_B T}\xi$ and use the fact that $R_A(t)=\sqrt{\beta\mu/2}\langle  A(t) \xi(0) \rangle$ \cite{cugliandolo1994off} we find the following familiar form of the FDT for Brownian suspensions 
\begin{equation}\label{eq:FDTbrownian}
\langle  A \rangle_{t}  - \langle  A \rangle_{0} =\beta h \int_0^{t} ds \langle  A(s) \dot{x}(0) \rangle_{0} \, .
\end{equation}
From this relation, it is straightforward to derive \emph{Green-Kubo expressions} for transport coefficients, such as the mobility and diffusivity. For instance, by choosing $A\equiv \dot{x}$ we find the well-known expressions
 \begin{equation}
 \mu=\lim_{t\to \infty}\frac{\langle  \dot{x} \rangle_t}{h}=\beta  \int_0^{\infty} ds \langle  \dot{x}(s) \dot{x}(0) \rangle_{0} ,\ \, D= \int_0^{\infty} ds \langle  \dot{x}(s) \dot{x}(0) \rangle_{0} \, .
\end{equation} 

\section{Non-equilibrium dynamics}\label{sec:4}
\subsection{Quantifying the violations of Detailed Balance}\label{4.1}

\begin{figure}
\centering
\includegraphics[width=\textwidth]{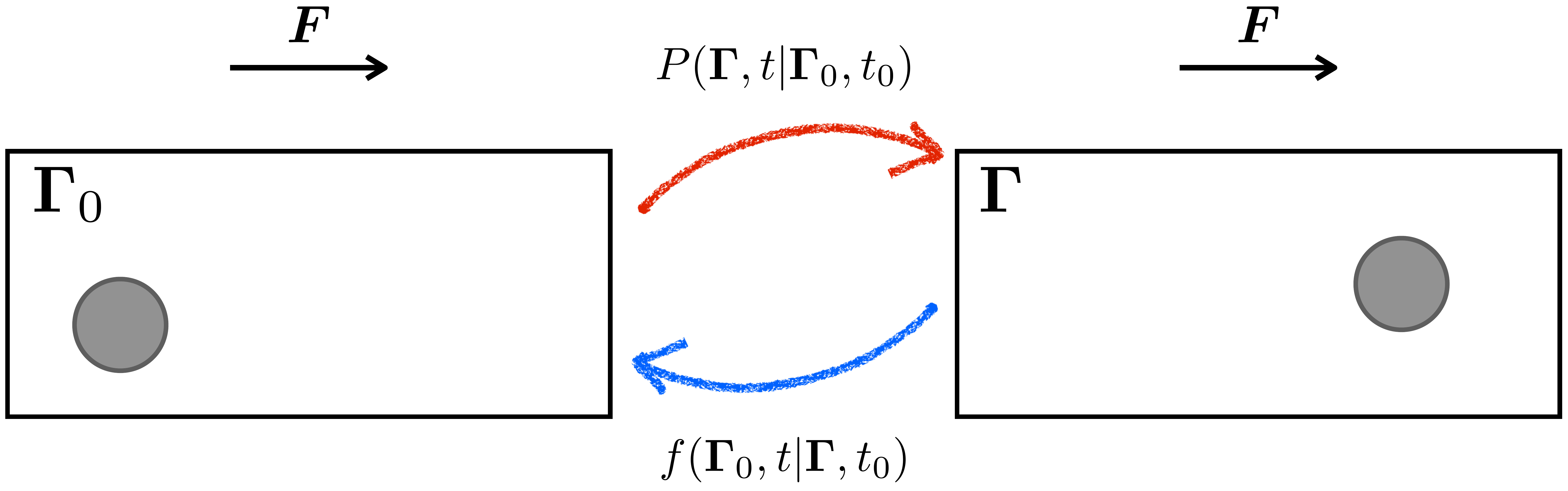}
\caption{ Illustration of the conditional probability $f$ introduced to quantify violations of DB. 
A force $\boldsymbol{F}$ is applied to a colloidal particle, which favors motion towards the right, and thus $f(\boldsymbol{\Gamma}_0, t |\boldsymbol{\Gamma}, t_0)<P(\boldsymbol{\Gamma}, t | \boldsymbol{\Gamma}_0\, t_0)$.  In the absence of external drive, DB is satisfied and there is no preference to move right or left, thus $f(\boldsymbol{\Gamma}_0, t |\boldsymbol{\Gamma}, t_0)=P(\boldsymbol{\Gamma}, t | \boldsymbol{\Gamma}_0\, t_0)$. 
}
\label{fig:sketch_f}
\end{figure}

In order to quantify the breakdown of DB we introduce the  conditional probability to go from $\boldsymbol{\Gamma}$ to $\boldsymbol{\Gamma}_0$ forward in time (see Fig.~\ref{fig:sketch_f}):
\begin{equation}\label{eq:defF}
f(\boldsymbol{\Gamma}_0, t | \boldsymbol{\Gamma}, t_0) \equiv \frac{\Psi_0(\boldsymbol{\varepsilon} \boldsymbol{\Gamma})}{\Psi_0(\boldsymbol{\Gamma}_0)} P(\boldsymbol{\varepsilon} \boldsymbol{\Gamma}_0, t | \boldsymbol{\varepsilon} \boldsymbol{\Gamma}, t_0) \, .
\end{equation}
 Whenever DB holds $f(\boldsymbol{\Gamma}_0, t |\boldsymbol{\Gamma}, t_0)=P(\boldsymbol{\Gamma}, t | \boldsymbol{\Gamma}_0\, t_0)$. 
When DB is violated, such identification is no longer valid. Nevertheless, it is still interesting to look at the evolution equation for $f(\boldsymbol{\Gamma}_0, t|\boldsymbol{\Gamma}, t_0)$ as it allows us to  quantify the breakdown of DB.
 Following \cite{PhysRevE.89.022127} we write the backward evolution equation for $P(\boldsymbol{\varepsilon} \boldsymbol{\Gamma}_0, t | \boldsymbol{\varepsilon} \boldsymbol{\Gamma}, t_0)$:
\begin{equation}
\fl \partial_t P(\boldsymbol{\varepsilon} \boldsymbol{\Gamma}_0, t | \boldsymbol{\varepsilon} \boldsymbol{\Gamma}, t_0) = \sum_i \varepsilon_i \mathcal{A}_i(\boldsymbol{\varepsilon} \boldsymbol{\Gamma}) \partial_i P(\boldsymbol{\varepsilon} \boldsymbol{\Gamma}_0, t | \boldsymbol{\varepsilon} \boldsymbol{\Gamma}, t_0) +D_i \partial_i^2 P(\boldsymbol{\varepsilon} \boldsymbol{\Gamma}_0, t | \boldsymbol{\varepsilon} \boldsymbol{\Gamma}, t_0) \, .
\end{equation}
If we multiply both sides of the latter equation by $\Psi_0(\boldsymbol{\varepsilon} \boldsymbol{\Gamma})/\Psi_0(\boldsymbol{\Gamma}_0)$, and use the definition of $f(\boldsymbol{\Gamma}_0, t|\boldsymbol{\Gamma}, t_0)$, we obtain
\begin{equation}\label{eq:backtoDB}
\partial_t f =  \Omega_0(\boldsymbol{\Gamma}) f + \sum_i\partial_i \left[ f \left( 2 \mathcal{A}_i^{\rm irr}(\boldsymbol{\Gamma}) - 2 D_i  \partial_i \log \Psi_0(\boldsymbol{\varepsilon}\boldsymbol{\Gamma}) \right] \right)
\end{equation}
where, for simplicity, we write $f(\boldsymbol{\Gamma}_0, t|\boldsymbol{\Gamma}, t_0)\equiv f$. 
 When DB holds,  we recover the Fokker-Planck equation: since the initial condition for $f(\boldsymbol{\Gamma}_0,t| \boldsymbol{\Gamma}, t_0) $ and $P(\boldsymbol{\Gamma}, t| \boldsymbol{\Gamma}_0, t_0)$ is the same by definition, it follows that $f= P(\boldsymbol{\Gamma}, t | \boldsymbol{\Gamma}_0, t_0)$, as expected.

The evolution of $f$ is generated by the  \emph{time-reversal operator,  $\bar{\Omega}_0$}, defined as
\begin{equation}\label{eq:forFcompact}
\fl \partial_t f (\boldsymbol{\Gamma}_0, t| \boldsymbol{\Gamma}, t_0) \equiv \bar{\Omega}_0(\boldsymbol{\Gamma}) f(\boldsymbol{\Gamma}_0, t| \boldsymbol{\Gamma}, t_0) = \Psi_0(\boldsymbol{\varepsilon} \boldsymbol{\Gamma}) \Omega_0^{\dagger}(\boldsymbol{\varepsilon}\boldsymbol{\Gamma}) \Big[ f(\boldsymbol{\Gamma}_0, t| \boldsymbol{\Gamma}, t_0) \Psi_0(\boldsymbol{\varepsilon}\boldsymbol{\Gamma})^{-1}\Big] \, .
\end{equation}
Its adjoint  
\begin{equation}\label{eq:TRonobservable}
\bar{\Omega}_0^{\dagger}(\boldsymbol{\Gamma}) = \Psi_0(\boldsymbol{\varepsilon}\boldsymbol{\Gamma})^{-1}\Omega_0(\boldsymbol{\varepsilon}\boldsymbol{\Gamma}) \Psi_0(\boldsymbol{\varepsilon}\boldsymbol{\Gamma}) \,,
\end{equation}
has a precise dynamical meaning: it evolves (forward in time) the observable $A$ along time-reversal paths, i.e.
\begin{equation}
\langle A \rangle_{(-t)} \equiv \int d\boldsymbol{\Gamma} \Psi_0(\boldsymbol{\Gamma}) e^{\bar{\Omega}^{\dagger} t} A(\boldsymbol{\Gamma}), \quad t >0
\end{equation}
generalizing the formulation of Baiesi and Maes \cite{BaiesiMaesNJP} to systems with odd variables. 
The above definition of the time-reversal operator and the expression of its adjoint in terms of the original generator, do not rely on any assumption concerning the steady-state distribution. From eq.~(\ref{eq:TRonobservable}), one easily finds the following  expression quantifying the breakdown of DB in terms of irreversible probability  fluxes:
\begin{equation}
\label{eq:beyondDB} \frac{\Omega_0^{\dagger}(\boldsymbol{\Gamma}) - \bar{\Omega}_0^{\dagger}(\boldsymbol{\Gamma})}{2}  =   \frac{\boldsymbol{\varepsilon} \boldsymbol{J}^{\rm irr}_0(\boldsymbol{\varepsilon} \boldsymbol{\Gamma})}{\Psi_0(\boldsymbol{\varepsilon}\boldsymbol{\Gamma})}  \cdot \boldsymbol{\nabla} \equiv \boldsymbol{\varepsilon}\boldsymbol{\mathcal{V}}^{\rm irr}(\boldsymbol{\varepsilon}\boldsymbol{\Gamma}) \cdot \boldsymbol{\nabla} \, .
\end{equation}
It immediately follows that DB  holds if and only if the time-reversal operator equals the original generator of the dynamics
\begin{equation}
\textrm{Detailed Balance} \quad  \Leftrightarrow \quad \Omega_0(\boldsymbol{\Gamma}) = \bar{\Omega}_0 (\boldsymbol{\Gamma}) \, .
\end{equation}
Since DB only constraints irreversible fluxes to vanish, it is thus expected that its breakdown only concerns their presence. 
The status of DB as a symmetry of the dynamics is now transparent.
These expressions are general and can  be exploited to quantify the breakdown of DB in systems with both odd and even variables, such as collections of interacting particles with inertia \cite{BaiesiIntertial} or noisy RLC circuits \cite{freitas2019stochastic}.

\subsection{General constraints on Non-Equilibrium Steady-States }\label{4.2}

It is useful to notice that the irreversible phase-space velocity quantifying the non-equilibrium character of the dynamics, or breakdown of DB,  is directly related to the housekeeping entropy production, i.e. the entropy being produced in a NESS. For a stochastic dynamics described by the Fokker-Planck equation (\ref{eq:genericFP}-\ref{eq:flux}), the housekeeping entropy production rate is  expressed as \cite{PhysRevE.85.051113}:
\begin{equation}\label{eq:housekeeping}
\langle \dot{S}_{{\rm hk}} \rangle_t=k_B\sum_i \int d\boldsymbol{\Gamma} D_i^{-1}\Psi(\boldsymbol{\Gamma}, t)\left( \boldsymbol{\varepsilon}\mathcal{V}^{\rm irr}_{i}(\boldsymbol{\varepsilon} \boldsymbol{\Gamma}) \right)^2 \geq 0
\end{equation}
where we recognize inside the integral the phase space velocity quantifying the violations of DB. The house-keeping term constitutes the entropy production necessary to sustain a NESS and it is positive as soon as DB is broken. It may be compared  with the standard formula  for the total entropy production \cite{PhysRevE.65.016102, PhysRevLett.95.040602,  PhysRevE.85.051113}:
\begin{equation}\label{eq:totalep}
\langle \dot{S}_{\rm tot} \rangle_t = k_B \sum_i \int d\boldsymbol{\Gamma}  D_i^{-1}  \Psi(\boldsymbol{\Gamma}, t)\left( \mathcal{V}^{\rm irr}_i(\boldsymbol{\Gamma}, t)\right)^2 \geq 0 \, .
\end{equation}
embedding contributions coming from the breakdown of DB as well as from any time-dependent processes (the so-called excess entropy production). By definition, in stationary conditions, $\dot{S}_{{\rm hk}}$ must reduce to the total entropy change $\dot{S}_{\rm tot}$,  since no time-dependent contribution is present.
While this is obvious for even variables only, in presence of odd variables  this observation provides a  \emph{general constraint on the phase-space velocity} such that:
\begin{equation}\label{eq:symmetryhk}
\left[ \mathcal{V}_{ i}^{\rm irr}(\boldsymbol{\varepsilon} \boldsymbol{\Gamma}) \right]^{2} = \left[ \mathcal{V}_{i}^{\rm irr}( \boldsymbol{\Gamma}) \right]^{2} \, .
\end{equation}
This constitutes one of our main results: as we will show below, this equality constraints any NESS  and allows to relate the nature of the  irreversible fluxes, responsible for the breakdown of DB, with the symmetry of the NESS distribution $\Psi_0$. 

In order to illustrate this general result and its consequences, we now consider a paradigmatic non-equilibrium model:  a one-dimensional particle in a periodic potential $U$, coupled to a thermal bath and driven out-of-equilibrium by a non-conservative force $F$  \footnote{Periodicity is needed in order to guarantee the existence of a steady-state.}. The Langevin equation governing the dynamics of such system is (see Fig. (\ref{fig:sketch_f}))
\begin{equation}\label{eq:particleonring}
\dot{x}(t)=p(t) \,\\ \dot{ {p}}(t)=-U'(x)+F-\gamma {p}(t)+\sqrt{2\gamma  k_BT}{\xi}(t) \, .
\end{equation} 
In this simple case,  the two  components  of irreversible part of the  phase-space velocity which enter in eq. \ref{eq:symmetryhk} read
\begin{eqnarray} 
\mathcal{V}_{x}^{\rm irr} (x,p)&=& 0 \\
\mathcal{V}_{p}^{\rm irr} (x,p)&=& -\gamma p  -\gamma k_BT \frac{\partial}{\partial p} \log \Psi_0(x,p) .
\end{eqnarray}
As we already mentioned, $\Psi_0$ does not necessary have a given parity under Time-Reversal. 
Nevertheless (without lack of generality) we may rewrite it as:
\begin{equation}\label{eq:factorPsi}
\Psi_0(x,p) = \mathcal{N} \exp (- \Phi (x,p)) = \mathcal{N} \exp\left[ -( \Phi_+ + \Phi_-)\right]
\end{equation}
where we have decomposed the generalized potential $\Phi$ into an even and odd contribution: 
$\Phi(x,-p)_{\pm}=\pm \Phi(x,p)$ \footnote{The generalized potential is assumed to be a differential function of the dynamic variables.}. Then,  the general constraint  eq.~(\ref{eq:symmetryhk})  yields
\begin{eqnarray}
 \Phi_+(x,p)&=&\beta\frac{p^2}{2}+\Lambda(x) \label{eq:contraint_phiplus}\\
 \mathcal{V}^{\rm irr}_{p}(x, p) &=&  \gamma k_B T \frac{\partial\Phi_{-}}{\partial p}\label{eq:contraint_phiminus}
\end{eqnarray}
where $\Lambda(x)$ is any derivable function of $x$.  Thus, the even part of the generalized potential  must  be of the latter form, meaning that even terms $p^{2n}$ with $n>1$ are strictly forbidden in the NESS distribution. As expected, the dependence of $\Psi(x, p)$ on the positions, which are associated to reversible fluxes, is not constrained  at all. 
Instead, irreversible fluxes are given by (gradients of) the odd part of  the generalized potential.  Thus,   eq.~(\ref{eq:symmetryhk})  establishes a  generic,  and explicit, relationship between irreversible fluxes and the symmetry of the NESS distribution.  

In the limit $U \to 0$,  the NESS distribution of this simple model can be computed analytically and consists in a 'tilted'  Gaussian distribution in $p$ and a  uniform distribution in $x$, such as :
\begin{equation}\label{eq:NESS_analytical}
\Psi_0(x, p) = \mathcal{N}\exp\left[-\frac{ \left(p -\langle p \rangle\right)^2}{2 k_B T} \right]
\end{equation}
where $\langle p \rangle = F/\gamma$ is the mean velocity of the particle. Combining eq.~(\ref{eq:NESS_analytical}) with eq.~(\ref{eq:contraint_phiminus}) we find $\Phi_{-} =- pF /  \gamma k_BT $ and consequently:
\begin{eqnarray}
\mathcal{V}_{p}^{\rm irr} &=& -  F  \\
\langle \dot{S}_{\rm hk} \rangle_0 &=& \frac{F^2}{\gamma T} > 0
\end{eqnarray}
which  reveals the non-equilibrium character of this system and the nature of its phase space velocity: it is proportional to the particle net current, resulting in a  positive house-keeping entropy production.

Now turning back into a more general discussion, eq.~(\ref{eq:symmetryhk}) can be rewritten in terms of the generalized potential as
\begin{equation}\label{eq:symmetryhk2}
\left[ \mathcal{A}_i^{\rm irr} + D_i\partial_i \Phi_{+} \right] D_i\partial_i\Phi_{-} =  0 \quad \forall i
\end{equation}
which, in order to be verified,  requires one of the two following set of conditions:
\begin{equation}
 \partial_i\Phi_{+}(\boldsymbol{\Gamma}) = - D_i^{-1}\mathcal{A}_i^{\rm irr} (\boldsymbol{\Gamma})  \quad   \mbox{or} \quad \partial_i \Phi_{-} = 0 \, .\label{eq:phioddzero}
\end{equation}
Note that if both constraints are fulfilled \emph{simultaneously} then the system obeys DB [see eq.~(\ref{eq:almostDB})]. 
Eq.~(\ref{eq:symmetryhk}) 'relaxes' one of the two constraints, allowing to distinguish two different ways of breaking DB, namely
\begin{eqnarray}
&&\Phi_{-} (\boldsymbol{\Gamma}) =0 \quad \mbox{but} \quad  \partial_i \Phi_{+} (\boldsymbol{\Gamma})  \neq -D_i^{-1}\mathcal{A}_i^{\rm irr} (\boldsymbol{\Gamma}) \label{eq:neqoverdamped}\\
&\mbox{or}& \nonumber \\
&&\partial_i \Phi_{+} (\boldsymbol{\Gamma})  =- D_i^{-1}\mathcal{A}_i^{\rm irr} (\boldsymbol{\Gamma}) \quad \mbox{but} \quad \Phi_{-} (\boldsymbol{\Gamma})  \neq 0 \, .\label{eq:nequnderdamped}
\end{eqnarray}
Any non-equilibrium system with only even variables falls into the first category. On the contrary,  systems with odd variables, such as the driven underdamped Brownian particle considered above or  the Active Ornstein-Ulhenbeck Particles that we treat in detail in section \ref{sec:5.2}, fall into the second category.
For  systems with odd variables, it follows that the irreversible flux is directly related to the   odd part of the NESS distribution, such that:
\begin{equation}\label{eq:phiodd_2}
\mathcal{V}_{i}^{\rm irr}(\boldsymbol{\Gamma}) = D_i\partial_i \Phi_{-} (\boldsymbol{\Gamma}) \, .
\end{equation}

\subsection{Extended Fluctuation-Dissipation Relations}\label{4.3}


We now consider a system initially prepared in a NESS with probability distribution $\Psi_{0}$. At time $t = 0$, we apply a perturbation such that  $\mathcal{A}\to \mathcal{A} + \delta \mathcal{A}$. 
Although the Agarwal FDR  eq.~(\ref{eq:agarwal}) remains valid, the lack of knowledge on $\Psi_0$  does not allow us to derive an explicit expression of the response function in terms of NESS correlations.
To move further,  an option would be to provide a reliable scheme to approximate $\Psi_0$. We follow this strategy in the next section when dealing with active particles. 
However, some general expressions can be establish through the NESS properties derived above. For instance, alternative FDRs can be written in a way that explicitly relate the non-equilibrium response of the system to the symmetry properties of $\Psi_0$. 

The Agarwal FDR can be rewritten in terms of the  even and odd part of the NESS distribution as
\begin{eqnarray}\label{eq:almost_extendedFDT}
\langle  A \rangle_{t}  - \langle  A \rangle_{0}& 
=& \sum_i \Big\{-\int_{0}^{t} ds\langle A(s) (\partial_i \delta \mathcal{A}_i) (0) \rangle_0 + \int_{0}^{t} ds \langle A(s) (\delta \mathcal{A}_i \partial_i \Phi_{+})(0) \rangle_0 \nonumber \\
&&+ \int_{0}^{t} ds \langle A(s) (\delta \mathcal{A}_i \partial_i \Phi_{-})(0) \rangle_0 \Big\} \, .
\end{eqnarray}
We  then make use of the relations, derived in the previous section, between the generalized potential and the phase space velocity to establish the following \emph{extended FDR}
\begin{eqnarray}\label{eq:neqsusceptibility2}
 \langle  A \rangle_{t}  - \langle  A \rangle_{0}&=& \sum_i \Big\{- \int_{0}^{t} ds\langle A(s) (\partial_i \delta \mathcal{A}_i) (0) \rangle_0 -D_i^{-1} \int_{0}^{t} ds \langle A(s) (\delta \mathcal{A}_i \mathcal{A}_i^{\rm irr})(0) \rangle_0 \nonumber \\&&+ D_i^{-1}\int_{0}^{t} ds \langle A(s) (\delta \mathcal{A}_i  \mathcal{V}_{i}^{\rm irr})(0)\rangle_0  \Big\}
\end{eqnarray}
which is  valid both for overdamped and underdamped systems in the presence any type of external perturbation $\delta \mathcal{A}$. As such, it generalizes the FDR appearing in \cite{SpeckSeifert, BaiesiMaesNJP, doi:10.1063/1.3669448}, derived  for an overdamped particle subjected to a conservative perturbation  $\delta \mathcal{A} \equiv - \mu \partial_x U$, for which it reduces to:
\begin{equation}\label{eq:neqsusceptibilityNJP}
\langle  A \rangle_{t}  - \langle  A \rangle_{0}= D_x^{-1}\mu \int_{0}^{t} ds \Bigg(   \left\langle A(s)  \dot{U}(0) \right\rangle_0 -  \left\langle  A(s) \partial_x U(0) \mathcal{V}^{\rm irr}_{x}(0)\right\rangle_0\Bigg) \nonumber 
\end{equation}
and if we consider the perturbation to be a constant force of amplitude $h$, $U \equiv  h x$, we end up with the non-equilibrium extension of the FDT eq.~(\ref{eq:FDTbrownian}) as derived by Speck and Seifert \cite{SpeckSeifert}:
\begin{equation}
\langle  A \rangle_{t}  - \langle  A \rangle_{0}= D_x^{-1} \mu h\int_{0}^{t} ds \Bigg(   \left\langle A(s)  \dot{x}(0) \right\rangle_0 - \left\langle  A(s) \mathcal{V}^{\rm irr}_{x}(0)\right\rangle_0\Bigg) . \nonumber 
\end{equation}
The  term comprising the irreversible phase space velocity is responsible for the non-equilibrium character of the dynamics as it encodes the breakdown of DB.
We notice by  comparing (\ref{eq:almost_extendedFDT}) and (\ref{eq:neqsusceptibility2})  that in the presence of odd variables,  full knowledge on $\Psi_0$ is not required to determine the response, but only on its odd part under Time-Reversal.

\section{Application to Active Particles}\label{sec:5}

\subsection{Active Brownian Particles}

\subsubsection{The model} We consider now $N$ two-dimensional overdamped Active Brownian Particles (ABP) moving in  two-dimensional space. They self-propelled with a constant velocity $v_0$,  along their orientation $\boldsymbol{n}_i=(\cos(\theta_i), \sin(\theta_i))$ and obey the following set of coupled Langevin equations
\begin{equation}\label{eq:ABP}
\dot{\boldsymbol{r}}_i(t)= \mu_0 \boldsymbol{F}_i+ v_0\boldsymbol{n}_i(t) + \boldsymbol{\xi}_i(t) \,,\ \, \dot{\theta}_i(t)= \nu_i(t)
\end{equation}
where $\boldsymbol{F}_i=-\partial U /\partial \boldsymbol{r}_i$, $\mu_0$ is the single particle mobility and $\boldsymbol{\xi}$ and $ \nu$ are zero-mean Gaussian noises  verifying $\langle  \boldsymbol{\xi}_i(t)  \boldsymbol{\xi}_j(t^{\prime})\rangle=2\mu_0k_BT\delta_{ij}\delta(t-t^{\prime})\bf{1}$ and $\langle \nu_i(t) \nu_j(t^{\prime})\rangle=2 D_{\theta} \delta_{ij}\delta(t-t^{\prime})$. 
It follows that 
\begin{equation} \label{ncorrelation}
\langle \boldsymbol{n}_i(t) \cdot \boldsymbol{n}_i(0) \rangle=e^{-D_{\theta}t} \ ,
\end{equation}
defining a persistence time 
$\tau=1 / D_{\theta}$. 
We define the P\' eclet number  {Pe}=$v_0/ \sigma D_{\theta}$
where $\sigma$ is a characteristic length-scale set, for instance, by the inter-particle potential.  Equilibrium is recovered both in the limit of $v_0 \to 0$ or $\tau\to0$. In both cases, the departure from equilibrium can be quantified by Pe. 
The generator, or Fokker-Planck operator, corresponding to this Langevin dynamics is 
\begin{equation}
\Omega_0 (\boldsymbol{\Gamma}) =\sum_i \frac{\partial}{\partial\boldsymbol{r}_i} ( \mu_0 k_B T \frac{\partial}{\partial \boldsymbol{r}_i} - \mu_0 \boldsymbol{F}_i - v_0 \boldsymbol{n}_i ) + D_{\theta} \sum_i \frac{\partial^2}{\partial \theta_i^2} 
\end{equation}
where $\boldsymbol{\Gamma}\equiv\{ \boldsymbol{r}_i,\theta_i\}$: All dynamical variables in ABP are considered even under Time-Reversal.

\subsubsection{Non-equilibrium character and non-interacting regime}

For ABP, DB is fullfilled if and only if
\begin{equation}
\left\{ 
  \begin{array}{ll}
  &\left( \mu_0 \boldsymbol{F}_i + v_0 \boldsymbol{n}_i \right)  \Psi_0 (\boldsymbol{\Gamma}) = \mu_0 k_B T \frac{\partial}{\partial \boldsymbol{r}_i} \Psi_0(\boldsymbol{\Gamma}) \nonumber \\
&\frac{\partial}{\partial \theta_i}  \Psi_0(\boldsymbol{\Gamma})= 0
 \end{array}\right.
\end{equation}
These two equations  can not simultaneously hold due to the self-propulsion term, and therefore, ABP generically break DB. By integrating the first condition, we get $\log \Psi_0 \sim  -\beta [U - \sum_i {v_0 \boldsymbol{n}_i \cdot \boldsymbol{r}_i}/{\mu_0}]$.  However, the second condition imposes $ \Psi_0$ to be a function of positions only, which is inconsistent with the first condition because of the term in $\boldsymbol{n}_i \cdot \boldsymbol{r}_i$. In the passive case, $v_0 \to 0$, DB is recovered together with the standard Boltzmann distribution.

An illustrative example for which we can explicitly compute the phase space velocity is a free ABP. The Fokker-Plank generator reads $\Omega_0 = \left[ {\partial}_{\boldsymbol{r}} ( \mu_0 k_BT \partial_{\boldsymbol{r}} - v_0 \boldsymbol{n}) + D_{\theta}  \frac{\partial^2}{\partial \theta^2} \right] $. A stationary solution of the Fokker-Planck equation can be derived  $\Psi_0(\boldsymbol{r},\theta)=\rho_0/2\pi$ \cite{EPJCates}. The phase space velocity corresponding to this homogeneous NESS is  
\begin{equation}
\boldsymbol{\mathcal{V}}^{\rm irr}(t)= ( \mathcal{V}_{\boldsymbol{r}}^{\rm irr}, \mathcal{V}_{\theta}^{\rm irr}) = (v_0 \boldsymbol{n}(t), 0)
\end{equation} 
Since the system  is overdamped, there are no reversible fluxes.  
In order to apply the extended FDR, we  consider a constant force perturbation $h$ applied along the $x$-axis.
By choosing $A= \dot{x}$ in eq.~(\ref{eq:neqsusceptibilityNJP}) we obtain 
\begin{equation}
\frac{\langle \dot{x}\rangle_t}{h}=\beta\int_0^t ds \langle \dot{x} (s) \dot{x} (0)\rangle_0-\beta v_0 \int_0^t ds \langle \dot{x} (s) \cos \theta(0)\rangle_0
\end{equation}
resulting in the following extended Stokes-Einstein relation in the long time regime
\begin{equation}
D/\mu_0= k_B T+ \frac{v_0^2}{2D_{\theta}\mu_0}\, .
\end{equation}
Therefore, non-interacting ABP fulfill the Stokes-Einstein relation with an effective temperature 
\begin{equation}
T_{\rm eff}/T= 1+\frac{v_0^2}{2D_{\theta}\mu_0 k_B T} \,  .
\end{equation} 
It is worth noting here that, although ABP generically break DB in a fundamental way (and do not allow for a zero current steady-state solution), in the  non-interacting limit they admit a  NESS which fulfills the Stokes-Einstein relation.

\subsubsection{Interacting regime: an effective Markovian description} As as evidenced by the extended FDR eq.~(\ref{eq:neqsusceptibility2}), 
the response of a non-equilibrium system is not completely determined by  NESS correlations of physical observables, but also depends on the specific form of its phase space velocity.
In other to  establish explicit FDR for interacting ABP one can approximate the dynamics by an effective equilibrium one that fulfills DB. Such kind of approximation has been used for ABP  and also AOUP and come under different names, the most usual ones being Unified Colored Noise  and Fox approximation \cite{HanggiRev, maggi2015, MarconiMaggi, Wittmann2017effective}. In both cases, the steady-state distribution does not correspond to the equilibrium Boltzmann distribution in terms of the energy function of the original dynamics,  but a `Boltzmann-like' distribution in terms of an effective energy function generating the approximated dynamics. Despite the non-Boltzmann character of the steady-state distribution resulting from these approaches, all the difficulties associated with the absence of DB are lifted and one can readily derive a FDR by direct application of the general results presented in the previous section. 

To be more specific, we turn now  into the analysis of interacting ABP within the Fox approximation \cite{Fox,Fox2}, as we previously presented in \cite{SaraPRL}. The starting point is to integrate out the angular variables appearing in the ABP dynamics. As usual, the integration of some stochastic variables introduces memory in the dynamics, here in the form of a colored noise with correlation time $\tau$. The equations of motion (\ref{eq:ABP}) can be approximated by 
\begin{equation}\label{eq:RABP}
\dot{\boldsymbol{r}}_i(t)= \mu_0 \boldsymbol{F}_i+ \boldsymbol{\eta}_i(t) 
\end{equation}
where the noise $\boldsymbol{\eta}_i$ is approximately Gaussian with zero mean and variance $\langle \boldsymbol{\eta}_i(t) \boldsymbol{\eta}_j(s) \rangle = (2 \mu_0k_BT  \delta(t-s)  + {v_0^2}e^{-  |t - s|/\tau}/2) \delta_{ij}\mathbf{1} $. 
The ABP dynamics in the reduced configuration space $\tilde{\boldsymbol{\Gamma}}\equiv\{\boldsymbol{r}_i\}$ is approximated by an effective  Fokker-Planck dynamics generated by the operator :
\begin{equation}
\Omega_0^{M}(\tilde{\boldsymbol{\Gamma}})  = \sum_{\alpha} \partial_{\alpha}   \left( \sum_{\beta} \partial_{\beta}   \mathcal{D}_{{\beta}\alpha}(\tilde{\boldsymbol{\Gamma}}) -\mu_0  {F}_{\alpha}(\tilde{\boldsymbol{\Gamma}}) \right)
\end{equation}
where  $\mathcal{D}\equiv\{ \mathcal{D}_{{\beta}\alpha} \}$ is an effective $2N \times 2N$ diffusivity tensor and where greek indices run over the spatial coordinates and the particle labels. To first order in $\tau \mu_0 \partial_{\alpha}F_{\beta} $, it reads
 \begin{equation}\label{FoxDF}
 \mathcal{D}_{{\alpha}{\beta}} (\tilde{\boldsymbol{\Gamma}})  = \mu_0 k_B T \delta_{{\alpha}{\beta}} + \frac{v_0^2 \tau}{2}\left( \delta_{{\alpha}{\beta}}+ \tau \mu_0 \partial_{\alpha} F_{\beta} \right) \, .
 \end{equation}
Note that the Fox approximation is meaningful only when $|\tau \mu_0 \partial_{\alpha} F_{\beta}| < 1$.
As we show below, this effective dynamics fulfills DB. 
The first step is to write the condition of DB eq.~(\ref{eq:gardinercondition}) for the stationary probability density $\Psi_0(\tilde{\boldsymbol{\Gamma}})$:
\begin{equation}\label{eq:effDtensor}
\sum_{\beta} \mathcal{D}_{{\beta}{\alpha}}(\tilde{\boldsymbol{\Gamma}}) \partial_{\beta} \Psi_0(\tilde{\boldsymbol{\Gamma}}) = \Psi_0(\tilde{\boldsymbol{\Gamma}}) \left( \mu_0 F_{\alpha} - \sum_{\beta} \partial_{\beta} \mathcal{D}_{{\beta}{\alpha}} (\tilde{\boldsymbol{\Gamma}}) \right) \, .
\end{equation}
We then multiply both  sides by $\mathcal{D}^{-1}_{{\alpha}{\gamma}}$ and sum over the index ${\alpha}$ to get:
\begin{equation}\label{eq:effforce}
 \partial_{\gamma} \log \Psi_0(\tilde{\boldsymbol{\Gamma}}) = \sum_{\alpha}\mathcal{D}^{-1}_{{\alpha}{\gamma}}(\tilde{\boldsymbol{\Gamma}}) \left[\mu_0 F_{\alpha} - \sum_{\beta}  \partial_{\beta} \mathcal{D}_{{\beta}{\alpha}}(\tilde{\boldsymbol{\Gamma}}) \right] \equiv \beta F_{\gamma}^{\rm eff}(\tilde{\boldsymbol{\Gamma}})
\end{equation}
Using  eq.~(\ref{FoxDF}) and eq.~(\ref{eq:effforce}) the effective force may we re-expressed as \cite{Wittmann2017effective}:
\begin{equation}\label{eq:effforce_curvature}
\beta F_{\gamma}^{\rm eff} (\tilde{\boldsymbol{\Gamma}}) = \frac{\mu_0}{D_a} F_{\gamma}(\tilde{\boldsymbol{\Gamma}}) - \left(\frac{\mu_0 v_0 \tau }{2 D_a }\right)^2 \sum_{\alpha} \partial_{\gamma} (F_{\alpha}(\tilde{\boldsymbol{\Gamma}}))^2 - \partial_{\gamma} \log[ \det \mathcal{D}(\tilde{\boldsymbol{\Gamma}})]
\end{equation}
where  $D_a \equiv \mu_0 k_B T + v_0^2 \tau /2$. It is now straightforward to verify that 
$\partial_{\beta} F_{\gamma}^{\rm eff} - \partial_{\gamma} F_{\beta}^{\rm eff} = 0 $. As a result, the system fulfills DB and therefore the FDT. This in turn implies that  $\boldsymbol{F}^{\rm eff} $ derives from an effective potential, such that an analytical expression for $\Psi_0$ in terms of an effective energy function can be derived \cite{MarconiMaggi}. 

Before leaving this section, it is worth mentioning that a diagonal-Laplacian approximation for $\mathcal{D}(\tilde{\boldsymbol{\Gamma}})$ was recently introduced and verified \textit{a posteriori} \cite{FarageBrader, Wittmann_2016, Rein_2016, Wittmann2017effective}. Within this approximation the effective force $\boldsymbol{F}_{i}^{\rm eff}$ on particle $i$ reads:
\begin{equation}
\boldsymbol{F}^{\rm eff}_{i}(\tilde{\boldsymbol{\Gamma}}) =k_BT ( \mu_0{\boldsymbol{F}_{i}} - {\partial_{i} \mathcal{D}_{i} })/{\mathcal{D}_{i}}
\end{equation}
with
\begin{equation}
\mathcal{D}_{i}(\tilde{\boldsymbol{\Gamma}}) = D_a \left( \frac{1}{ 1- \tau\mu_0 \partial_i \cdot \boldsymbol{F}_i}\right) 
\end{equation}
where $\partial_i \cdot \boldsymbol{F}_i = \partial_{x_i} F_i^{x} + \partial_{y_i} F^{y}_i$, further simplifying the analysis of ABP within the Fox approximation.

\subsection{Active Ornstein-Uhlenbeck Particles}\label{sec:5.2}

\subsubsection{The model}\label{aoupmodel}
We consider in this section a similar model of self-propelled particles, now governed  by the following set of two-dimensional overdamped Langevin equations 
\begin{equation}\label{eq:activelangevin}
\dot{\boldsymbol{r}}_i(t) = \mu_0 \boldsymbol{F}_i + \boldsymbol{v}_{i}
\end{equation}
\begin{equation}\label{eq:ornsteinuhlenbeck}
\dot{\boldsymbol{v}}_{i}(t) = - \frac{ \boldsymbol{v}_{i}}{\tau} +\sqrt{\frac{2D_0}{\tau^2}} \boldsymbol{\eta}_i(t)
\end{equation}
where $\boldsymbol{F}_i \equiv - \partial U/\partial{{\boldsymbol{r}}_i}$ is a conservative force acting on particle $i$ (whose origin can be  interactions with other particles or an external potential) and $\boldsymbol{v}_{i}$ is the fluctuating  self-propulsion velocity which is described by an Ornstein-Uhlenbeck process with characteristic persistence time $\tau$. Self-propulsion   introduces persistence in the spatio-temporal dynamics of the active particles via the autocorrelation function of the self-propulsion velocity $\langle \boldsymbol{v}_{i}(t) \boldsymbol{v}_{j}(t') \rangle = D_0/ \tau e^{-|t - t'|/\tau} \delta_{ij} \boldsymbol{1}$ and reduces to passive (equilibrium) Brownian motion in the limit $\tau \to 0$, for which $\langle\boldsymbol{v}_{i}(t) \boldsymbol{v}_{j}(t') \rangle\to 2D_0\delta(t-t')\delta_{ij}\boldsymbol{1}$. Although a standard  thermal noise could be added into the Langevin equation of AOUP, such contribution is assumed to be small with respect to the active noise $\boldsymbol{v}$ and might be considered redundant, as it is not needed to recover equilibrium.
These AOUP can be thought of as an approximate treatment of the ABP dynamics. Indeed, the reduced ABP dynamics obtained from the integration of the angular variables eq.~(\ref{eq:RABP}) can be identified, in the absence of translational noise ($T=0$), to the AOUP dynamics by setting $v_0^2/2$ (in ABP) to $D_0/\tau$ (in AOUP).

Although originally thought of as an overdamped process, eq.~(\ref{eq:activelangevin}) involves velocity variables $\boldsymbol{v}$ that can be   considered as being odd under Time-Reversal. 
Following this interpretation, eq.~(\ref{eq:activelangevin}) can be  rewritten as an underdamped Langevin process \cite{FodorPRL}
\begin{eqnarray}\label{eq:AOUPfrench}
\dot{\boldsymbol{r}}_i(t)& =& \boldsymbol{p}_i \\
\dot{\boldsymbol{p}}_i(t) &=& \mu_0 (\sum_j \boldsymbol{p}_j \cdot \partial_j) \boldsymbol{F}_i - \frac{\boldsymbol{p}_i}{\tau} + \mu_0\frac{\boldsymbol{F}_i}{\tau} + \sqrt{\frac{2 D_0}{\tau^2}} \boldsymbol{\eta}_i(t) \label{eq:momenta1}
\end{eqnarray}
where $\partial_i \equiv \partial / \partial \boldsymbol{r}_i$. 
The corresponding generator reads
\begin{equation}\label{eq:FPforAOUP}
\Omega_0  (\boldsymbol{\Gamma})= \sum_i\left[- \boldsymbol{p}_i\cdot \partial_i - \partial_{\boldsymbol{p}_i} \cdot \left( \mu_0 (\sum_j \boldsymbol{p}_j \cdot \partial_j)\boldsymbol{F}_i- \frac{\boldsymbol{p}_i}{\tau } + \mu_0 \frac{\boldsymbol{F}_i}{\tau} - \frac{D_0}{\tau^2} \partial_{\boldsymbol{p}_i} \right) \right]
\end{equation}
where $\partial_{\boldsymbol{p}_i}  \equiv \partial / \partial {\boldsymbol{p}_i} $ and $\boldsymbol{\Gamma}=\{\boldsymbol{r}_i,\boldsymbol{p}_i\}$.
\subsubsection{Effective equilibrium regime}\label{aoupeq}

For  the AOUP model, the DB condition eq.~(\ref{eq:gardinercondition}) reduces to:
\begin{equation}\label{eq:GardinerforAOUP}
 \frac{1}{\tau^2} \partial_{\boldsymbol{p}_i} \log \Psi_0(\boldsymbol{\Gamma}) =\beta( \sum_j\boldsymbol{p}_j \cdot \partial_j) \boldsymbol{F}_i - \frac{\boldsymbol{p}_i}{D_0\tau}
\end{equation}
Its formal solution can be expressed up to a  function $\Lambda(\{\boldsymbol{r}_i\})$ that only depend on space variables such as 
\begin{equation}\label{eq:equilibriumPsi}
\Psi_0  = \exp \left[- \Lambda(\{\boldsymbol{r}_i\}) - \frac{\beta \tau^2}{2} (\sum_i \boldsymbol{p}_i \cdot \partial_i)^2 U - \sum_i \frac{\tau}{D_0} \frac{\boldsymbol{p}_i^2}{2} \right]
\end{equation}
where $\beta\equiv \mu_0 /D_0$.  In order for DB to hold, the system must fulfill the following equation
\begin{eqnarray}\label{eq:DBcondition2}
\sum_i \left[ \partial_i \Lambda +\frac{\beta \tau^2}{2} (\sum_j \boldsymbol{p}_j \cdot \partial_j)^2 \partial_i U- \frac{\beta^2 D_0 \tau}{2} \partial_i \sum_j|\partial_j U |^2 - \beta \partial_i U \right] \Psi_0 = 0 . \nonumber \\
\end{eqnarray}
This equation does not have a solution because of the second term  comprising a $p-$dependence. Interestingly, for a  potential with vanishing third derivatives the latter term vanishes and an exact `equilibrium' solution exists  \cite{FodorPRL, Bonilla}:
\begin{equation}\label{eq:equilibriumPsi2}
\Psi_{\rm eq} = \mathcal{N}\exp \left[ -\beta U  -  \frac{\tau}{2}\sum_i \left( \frac{\boldsymbol{p}_i^2}{D_0} +\beta^2 D_0  |\partial_i U|^2 \right) - \frac{\beta \tau^2}{2} (\sum_i \boldsymbol{p}_i \cdot \partial_i)^2 U \right] \, .
\end{equation}
We wrote equilibrium in quotes because, contrarily to the standard Boltzmann measure, the probability of a given configuration is not solely given by $e^{-\beta U}$, but by a more complicated function, also involving $\boldsymbol{p}$. This form is not {\it a priori} obvious from the mere inspection of the generator of the microscopic dynamics. [The same remark holds  for ABP within the Fox approximation discussed earlier, as the effective potential eq. \ref{eq:effforce_curvature} can hardly be guessed from the original Fokker-Planck equation.] However, in this equilibrium-like regime, AOUP fulfill DB, there are no irreversible fluxes, and  the FDT holds. 
An equilibrium solution also exists in the case of non-interacting particles $U =0$ for which AOUP are formally equivalent to an ideal  gas of underdamped particles. In all other cases, for a generic $U$, the model breaks DB and therefore falls out-of-equilibrium.

\subsubsection{Non-equilibrium regime: Chapman-Enskog expansion}\label{5.2.2}
An approximated stationary distribution $\Psi_0$ for AOUP, beyond its equilibrium-like regime, has recently been derived via the Chapman-Enskog expansion by Bonilla \cite{Bonilla}.
Our aim being to study the impact of activity on the response of an interacting system, we briefly present the  Chapman-Enskog results as appeared  in \cite{Bonilla} and use them to establish extended FDR for AOUP. 

The   Chapman-Enskog expansion constitutes a standard perturbative approach to derive the Navier-Stokes equation from the Boltzmann equation \cite{Grad1958, Soto2016KineticTA}. It is based on the notions of local equilibrium and time scale separation. The latter is accounted for by the introduction of a small parameter $\epsilon = \ell /L$ defined as the ratio between a microscopic and a macroscopic characteristic length. In kinetic theory, $\ell$ is typically the \textit{mean free path} between collisional events and $L$  the size of the system. 
Likewise, we may associate to the AOUP  two different scales: a microscopic one associated to the persistence time $\tau$ and diffusive length $\sqrt{D_0 \tau}$ (characterizing the local persistence due to activity),  and a mesoscopic one associated to the inter-particle interactions, with a 'slow' characteristic time $\tau_0$ and a large characteristic length $L$.  We introduce the ratio parameter  $\epsilon \equiv \frac{\sqrt{D_0 \tau}}{\mathit{L}} \equiv \frac{\tau}{\tau_0} $  and rescale the AOUP equations of motion Eqs. (\ref{eq:AOUPfrench}-\ref{eq:momenta1}) according to $t\equiv t /\tau_0 $, $\boldsymbol{r} \equiv \boldsymbol{r}/ \mathit{L}$ and $\boldsymbol{F} \equiv \boldsymbol{F}\mathit{L} / \beta^{-1} $. The Fokker-Planck equation of AOUP can thus be written in the following non-dimensional form
\begin{equation}\label{eq:adimentionalFP}
\sum_i \partial_{\boldsymbol{p}_i} \cdot \left(\boldsymbol{p}_i + \partial_{\boldsymbol{p}_i} \right) \Psi = \epsilon \partial_t  \Psi + \epsilon\sum_i [ \boldsymbol{p}_i  \cdot \partial_i + \boldsymbol{F}_i  \cdot \partial_{\boldsymbol{p}_i} + \epsilon \partial_{\boldsymbol{p}_i}  \cdot (\sum_j \boldsymbol{p}_j\cdot \partial_j) \boldsymbol{F}_i]  \Psi 
\end{equation}
From here, the idea is to carry on a perturbative expansion in $\epsilon$.
For $\epsilon = 0 $, a solution of eq.~(\ref{eq:adimentionalFP}) is:
\begin{equation}\label{eq:solutionepszero}
\Psi^{(\epsilon=0)} (\boldsymbol{\Gamma})= \frac{e^{- \sum_i \boldsymbol{p}_i^2/2}}{(2 \pi)^{N}} R(\boldsymbol{r}, t)
\end{equation}
where $R(\boldsymbol{r}, t)$ is the normalized marginal density such that $\int \Pi_id\boldsymbol{r}_i R(\boldsymbol{r}, t) = 1$. For a system with strong time-scale separation, $\epsilon \ll 1 $,  we assume that the functional dependence on $\{\boldsymbol{p}_i\}$ and $R$ in eq.~(\ref{eq:solutionepszero}) is preserved, and expand the probability distribution as a power series in $\epsilon$:
\begin{equation}\label{eq:CEansatz}
\Psi(\boldsymbol{\Gamma})= \frac{e^{- \sum_i \boldsymbol{p}_i^2/2}}{(2 \pi)^{N}} R(\boldsymbol{r}, t; \epsilon) + \sum_j \epsilon^j \phi^{(j)} (\boldsymbol{\Gamma}, R) \, .
\end{equation}
The crucial assumption of the Chapman-Enskog method is to still interpret $R$ in (\ref{eq:CEansatz}) as the marginal distribution embedding the spatiotemporal dependence upon integration over the velocities. This corresponds to impose for $\phi^{(j)}$:
\begin{equation}\label{eq:consistencyequation}
\int \Pi_i d\boldsymbol{p}_i \; \phi^{(j)}(\boldsymbol{\Gamma}, R) = 0 \quad \forall j \, .
\end{equation}
The \textit{ansatz} eq.~(\ref{eq:CEansatz}) is inserted into eq.~(\ref{eq:adimentionalFP}), resulting in a hierarchy of equations for the various terms in the expansion. We solve the set of equations up to $\sim o(\epsilon^3)$ and  obtain the following (now made dimensional) probability distribution \cite{FodorPRL,Bonilla}:
\begin{eqnarray}\label{eq:CEpsi}
\Psi_0 (\boldsymbol{\Gamma})&\simeq& \mathcal{N} \exp \Bigg[ -\beta U - \sum_i \frac{\tau}{2} \left(\frac{\boldsymbol{p}_i^2}{D_0} + \beta^2 D_0 (\partial_i U)^2 - 3 \beta D_0 \partial_i^2 U \right)\nonumber  \\ 
&-& \frac{\tau^2}{2} \left( \beta (\sum_j \boldsymbol{p}_j \cdot \partial_j)^2 U +\beta D_0 \sum_{i, j} ( \boldsymbol{p}_j \cdot \partial_j) \partial_i^2 U \right) + \frac{\tau^3}{6} \beta (\sum_j \boldsymbol{p}_j \cdot \partial_j)^3 U \Bigg] \, . \nonumber \\
\end{eqnarray}
Some details on the derivation of eq.~(\ref{eq:CEpsi}) are given in the Appendix for the one dimensional case. The generalization to higher dimensions is straightforward but lengthy (see also \cite{Bonilla} for further details). 

We are now in the position of computing the  non-equilibrium response of AOUP up to third order in $\epsilon$. 
First, let us decompose $\Psi_0$ into its  symmetric and antisymmetric parts:
\begin{eqnarray}
\Phi_{+} &=& \beta U + \sum_i \frac{\tau}{2} \left(\frac{\boldsymbol{p}_i^2}{D_0} + \beta^2 D_0 (\partial_i U)^2 - 3 \mu_0 \partial_i^2U\right)+ \frac{\tau^2}{2} \beta (\sum_j\boldsymbol{p}_j \cdot \partial_j )^{2} U \nonumber \\\label{eq:AOUPphiplus} \\
\Phi_{-} &=& \sum_{i,j} \frac{\tau^2}{2} \beta D_0 ( \boldsymbol{p}_j\cdot \partial_j) \partial^2_i U - \frac{\tau^3}{6}\beta (\sum_i\boldsymbol{p}_i \cdot \partial_i)^{3} U \, . \label{eq:AOUPphiminus}
\end{eqnarray}
Note that, indeed $\Phi_+$ does not contain powers of $p$ larger than 2, as forbidden by the non-equilibrium constraint of eq.~ (\ref{eq:symmetryhk}). 
The signature  of the departure from equilibrium is all embedded in $\Phi_{-}$, being different from zero (see eq.~(\ref{eq:nequnderdamped})). Particularly, the latter vanishes for a quadratic potential, as expected from the  discussion in the previous section. 

We now apply a constant force $h$ along the $x$-axis on a tagged particle $n$.
In this case, the extended FDR  eqs.~(\ref{eq:almost_extendedFDT}-\ref{eq:neqsusceptibility2}) reads
\begin{equation}
 \langle A\rangle_t-  \langle A\rangle_0  =\delta \mathcal{A}_{p_n} \left[-(D_{p_n})^{-1} \int_0^{t} ds \langle A(s) \mathcal{A}_{p_n}^{\rm irr} (0) \rangle_0 + \int_0^{t} ds \langle A(s) \frac{\partial  \Phi_{-}}{\partial {{p}^x_n}}(0) \rangle_0 \right] \nonumber  \\
 \end{equation}
 where $\delta \mathcal{A}_{p_n} = \mu_0 {h}/\tau$ and $ D_{p_n} = D_0/\tau^2$, leading to
 \begin{equation}
 \langle A\rangle_t-  \langle A\rangle_0= {h} \left[-\tau \beta \int_0^{t} ds \langle A(s) \mathcal{A}_{p_n}^{\rm irr} (0) \rangle_0 + \frac{\mu_0}{\tau} \int_0^{t} ds \langle A(s) \frac{\partial  \Phi_{-}}{\partial {{p}^x_n}}(0)\rangle_0 \right] \, . \label{eq:neqresponse_almost}
\end{equation}
The first term is directly determined upon making the identification  $\mathcal{A}^{\rm irr}_{p_n} = \mu_0 (\sum_j \boldsymbol{p}_j \cdot \partial_j) {F}^x_n- {{p}^x_n}/{\tau}$. The  second integral requires the knowledge of the odd-symmetric part of $\Psi_0$, which, to third order in $\epsilon$ is given by eq.~(\ref{eq:AOUPphiminus}).  All in all, we derive the following FDR for AOUP
\begin{eqnarray}\label{eq:FDRnoneq}
\fl  \langle A\rangle_t-  \langle A\rangle_0 &=&\beta {h}  \Bigg[ \int_0^t ds \langle A(s){p}^x_n(0)  \rangle_0 - \tau \mu_0 \int_0^t ds \langle A(s)	(\sum_j  \boldsymbol{p}_j \cdot \partial_j) {F}^x_n(0) \rangle_0 \\ \fl &-&  \frac{1}{2} \mu_0 \tau D_0 \left( \int_0^t ds \langle A(s)\frac{\partial}{\partial x_n} ( \sum_j  \partial_j \cdot \boldsymbol{F}_j)(0) \rangle_0 - \frac{\tau}{D_0} \int_0^t ds \langle A(s)(\sum_j\boldsymbol{p}_j \cdot \partial_j)^2 {F}^x_n(0) \rangle_0 \right) \Bigg] \nonumber
\end{eqnarray}
By choosing $A \equiv \boldsymbol{p}_n$ we eventually obtain an extended Stokes-Einstein relation:
\begin{eqnarray}\label{eq:GKAOUP}
\fl \mu &=& \beta \Bigg[ D - \tau \mu_0 \int_0^{\infty} ds \langle {p}^x_n(s) ( \sum_j \boldsymbol{p}_j \cdot \partial_j) {F}^x_n(0)  \rangle_0 \\ \fl &-& \frac{1}{2}  \mu_0 \tau D_0 \left(\int_0^{\infty} ds \langle  {p}^x_n(s)  \frac{\partial}{\partial x_n} ( \sum_j \partial_j \cdot \boldsymbol{F}_j)(0)  \rangle_0 -\frac{\tau}{D_0} \int_0^{\infty} ds \langle {p}^x_n(s) ( \sum_j \boldsymbol{p}_j \cdot \partial_j)^2\boldsymbol{F}_n(0) \rangle_0 \right) \Bigg] \nonumber
\end{eqnarray}
where $D$ is the many-body diffusivity $D=\int_0^{\infty}ds \langle p_n^x(s)p_n^x(0)\rangle$. The expression above embeds the violations to the usual Stokes-Einstein relation due to the interplay between activity and inter-particle interactions, up to order $o(\tau^2)$. In the absence of interactions  the Stokes-Einstein relation is restored.  This is also true for the  case of a harmonic potential $U(r) =  k |\boldsymbol{r}|^2/2$, for which we are left with
\begin{equation}\label{eq:SEharmonic}
\mu= \beta^{\rm eff} \int_0^{t} ds \langle {p}^x_n (0) {p}^x_n(s)\rangle_0
\end{equation}
being $\beta^{\rm eff} = \beta ( 1 + \mu_0 \tau k)$ an effective temperature which depends on the stiffness of the external potential \cite{szamel2014self}. 
In contrast with ABP, the existence of a Stokes-Einstein relation for a harmonic potential in AOUP  is due to the fact that the  model fulfills DB. 

\section{Conclusions}\label{sec:6}
In this paper we have recalled, and discussed in detail, the pivotal role played by Detailed Balance as the defining feature of  equilibrium dynamics, and how its breakdown out-of-equilibrium can be quantified by the presence of irreversible steady-state fluxes. We have analyzed the symmetry properties of such fluxes under Time-Reversal for systems with odd and even variables, taking as  illustrative examples Langevin processes describing Brownian particles  both in the underdamped and overdamped regimes. We have developed a general formalism based on Fokker-Planck operators that allows us to express irreversible steady-fluxes in terms of the difference between the generator of the time-reversed dynamics and the original one.  

By making  the connection between the breakdown of Detailed Balance and  the different contributions to the entropy production, we derived a constraint of the irreversible steady-state fluxes. This general result   applies to non-equilibrium  systems and provides non-trivial  information for systems with dynamic variables which are odd under Time-Reversal. In particular, it constraints the functional dependence  of the NESS distribution on its odd variables. We then considered the linear response of a system in a NESS and derived extended Fluctuation-Dissipation Relations, allowing to express non-equilibrium response functions as NESS correlations and shedding light upon the nature of the different terms responsible for violations of the equilibrium Fluctuation-Dissipation Theorem. 

We then apply these general results and formalism to Active Brownian Particles (ABP) and Active Ornstein-Uhlenbeck Particles (AOUP). While ABP generically break Detailed Balance, AOUP fulfill Detailed Balance in the dilute limit, or in the case of a harmonic potential. The non-equilibrium nature of these two model systems is therefore not equivalent. We then analyze their linear response  in the presence of generic many-body interactions in an approximated fashion. In the case of ABP, we recall the Markov approximation method due to Fox and show that the effective dynamics resulting from it fulfills Detailed Balance, and therefore also the standard Fluctuation-Dissipation Theorem. For AOUP we exploit the Chapman-Enskog expansion performed in \cite{Bonilla}   which allows to derive an extended Fluctuation-Dissipation Relation beyond its effective equilibrium regime. We discuss the violations of the Stokes-Einstein relation in these models of active particles and show the possibility of quantifying them in terms of effective temperatures. 


Although some of the results presented here were known, as the existence of an effective equilibrium regime of AOUP and the NESS solution obtained from the Chapman-Enskog expansion, the discussion about the violations of DB and the FDT were scattered and scarce. The extended FDR, as well as the connection with the parity of the NESS distribution and the constraint on the phase-space velocity we derived, enrich previous discussions on non-equilibrium response, clarify the non-equilibrium nature of  active model systems and provide a set of analytic results that should be of interest to study non-equilibrium systems in general, well beyond the context of active systems.

\section*{Acknowledgments}
We warmly thank Jorge Kurchan, Matteo Polettini and Patrick Pietzonka for useful discussions and suggestions.  
S.D.C. acknowledges funding from the European Union's Horizon 2020 Framework Programme/European Training Programme 674979 €"NanoTRANS.
D.L.  acknowledges MCIU/AEI/FEDER for financial support under grant agreement RTI2018-099032-J-I00.
I.P. acknowledges support from Ministerio de Ciencia, Innovaci\'on y Universidades (Grant No. PGC2018-098373-B-100 AEI/FEDER-EU) and from
Generalitat de Catalunya under project 2017SGR-884, and Swiss National Science Foundation Project No. 200021-175719.

\section*{References}
\bibliographystyle{unsrt}

\bibliography{biblio.bib}

\newpage
\section*{Appendix A: Chapman-Enskog method for one dimensional AOUP}
Here we report some details of the Chapman-Enskog procedure for a one dimensional system  of AOUP. The starting point is the evolution equation (\ref{eq:adimentionalFP}) together with  the ansatz of Eqs.~(\ref{eq:CEansatz}-\ref{eq:consistencyequation}).
By integrating (\ref{eq:adimentionalFP}) over $p$ we obtain a conservation equation for $R$:
\begin{equation}\label{eq:conservativeqR}
\partial_t R = - \partial_x \sum_j J^{(j)} =  - \sum_j \epsilon^j \partial_x \int dp \; p \; \phi^{(j)}  \equiv \sum_j \epsilon^j \mathcal{F}^{(j)}
\end{equation}
where $J = \sum_j J^{(j)}$ is the probability flux.
\\We now substitute (\ref{eq:conservativeqR}) into (\ref{eq:adimentionalFP}) and we obtain to first order in $\epsilon$ an equation for $\phi^{(1)}$:
\begin{equation}\label{eq:epsilon1}
\mathcal{O}(\epsilon): \mathcal{L}[\phi^{(1)}] \equiv \partial_p( p \; \phi^{(1)} + \partial_p \phi^{(1)} ) = p \frac{e^{-p^2/2}}{\sqrt{2 \pi}} (\partial_x R +  U' R)
\end{equation}
In eq.~(\ref{eq:epsilon1}) $\phi^{(1)}$ has a gaussian dependence in momenta and we find:
\begin{equation}\label{eq:Phi1}
\phi^{(1)} = -p \frac{e^{-p^2/2}}{\sqrt{2 \pi}} \mathcal{D}R
\end{equation}
with $\mathcal{D}R \equiv U' R + \partial_x R$. Note that $\mathcal{L}[ e^{-p^2/2}] =0$ such that the functions $\phi^{(j)}$ are defined up to a constant in $p$ which is  uniquely determined by conditions (\ref{eq:consistencyequation}). 
The second order in $\epsilon$ reads:
\begin{equation}\label{eq:epsilon2}
\mathcal{O}(\epsilon^2) : \mathcal{L}[\phi^{(2)}] = \frac{e^{-p^2/2}}{\sqrt{2 \pi}} [ \mathcal{F}^{(1)} (R) + (p^2 -1) U'' R ] + p \partial_x \phi^{(1)} - U' \partial_p \phi^{(1)}
\end{equation}
From eq.~(\ref{eq:Phi1}) and (\ref{eq:epsilon2}) it follows that $\phi^{(2)}$ is even in p, such that $\mathcal{F}^{(2)} = \int dp \, p \, \phi^{(2)} = 0$.
\\ Using 
\begin{equation}
\mathcal{L}\left[p^2 \frac{e^{-p^2/2}}{\sqrt{2 \pi}}\right] = -2(p^2 -1) \frac{e^{-p^2/2}}{\sqrt{2 \pi}}
\end{equation}
 we find:
\begin{equation}\label{eq:Phi2}
\phi^{(2)} = \alpha(x) (p^2 -1) \frac{e^{-p^2/2}}{\sqrt{2 \pi}}
\end{equation}
with $\alpha = \frac{1}{2} (\partial_x \mathcal{D}R + U' \mathcal{D}R - U'' R )$.
\\
To third order in $\epsilon$ the evolution equation gives:
\begin{equation}
\mathcal{O}(\epsilon^3): \mathcal{L}[\phi^{(3)}] = \frac{\delta \phi^{(1)}}{\delta R} \mathcal{F}^{(1)}(R) + p \partial_x \phi^{(2)} -U' \partial_p \phi^{(2)} - U'' \partial_p(p \phi^{(1)})
\end{equation}
which, using (\ref{eq:Phi1}) and (\ref{eq:Phi2}), together with the identity:
\begin{equation}
\mathcal{L} \left[ p^3 \frac{e^{-p^2/2}}{\sqrt{2 \pi}} \right] = \frac{e^{-p^2/2}}{\sqrt{2 \pi}}3 p (2 -p^2)
\end{equation}
 gives:
\begin{equation}\label{eq:Phi3}
\phi^{(3)} = p \frac{e^{-p^2/2}}{\sqrt{2 \pi}} (\beta(x) p^2 + \gamma(x))
\end{equation}
with $\beta = -\frac{1}{3} (\mathcal{D}\alpha- U'' \mathcal{D}R)$ and $\gamma = \mathcal{D}\partial_x \mathcal{D}R -\partial_x \alpha + U' \alpha$.
We truncate the hierarchy of equations to $\sim o(\epsilon^3)$ and, going back to the conservative equation (\ref{eq:conservativeqR}), we find an expression for the first three components of the flux as a function of $R$:
\begin{eqnarray}\
J^{(1)} &=& -\mathcal{D}R \\
J^{(2)} &=& 0 \\
J^{(3)} &=& \partial_x U'' R
\end{eqnarray}
We have therefore a closed approximated equation for the evolution of  $R$:
\begin{equation}\label{eq:approxforR}
\frac{\partial R}{\partial t} =- \epsilon \partial_x [ - \partial_x R - U' R + \epsilon^2 \partial_x (U'' R) ] + o(\epsilon^3)
\end{equation}
Eq.~(\ref{eq:approxforR}) is readily solved by looking at the zero-flux steady-state with solution:
\begin{equation}\label{eq:explicitR}
R(x; \epsilon) = \exp\left[-U - \epsilon^2 \left(\frac{1}{2} U'^2 - U''\right) \right] + o(\epsilon^3)
\end{equation}
In turn (\ref{eq:explicitR}) is used to find the explicit expressions of $\phi^{1, 2, 3}$ which are substituted in (\ref{eq:CEansatz}) to get:
\begin{equation}\label{eq:Psi1}
\Psi_0(x, p; \epsilon) = \Psi^{(0)} \left[ 1 + \frac{\epsilon^2}{2}(1-p^2) U'' + \epsilon^3 \left( \frac{p^3}{6}U''' - \frac{p}{2} U''' \right) \right] +o(\epsilon^3)
\end{equation}
Finally, we use (\ref{eq:explicitR}) into (\ref{eq:Psi1}) and  interpret the term in square brakets as the linearization of an exponential $\exp(x) \sim 1 + x$ so that:
\begin{equation}\label{eq:Psi1D}
\fl \Psi_0(x, p;\epsilon) \sim \exp \left[ -\frac{p^2}{2} -U  -\frac{\epsilon^2}{2}\left( p^2 U'' + U'^2 - 3U'' \right) + \frac{\epsilon^3}{2} \left( \frac{p^3}{3} U''' - p U''' \right) \right] + o(\epsilon^3) 
\end{equation}

\end{document}